\documentclass[12pt]{article}

\usepackage{amsmath,amsfonts,epsfig,color,latexsym,cite}

\newcommand{\be}{\begin{equation}}
\newcommand{\ee}{\end{equation}}
\newcommand{\bea}{\begin{eqnarray}}
\newcommand{\eea}{\end{eqnarray}}

\newcommand{\SYM}{{\rm super-Yang-Mills}}

\def\yop{ y_1^{\prime} } 
\def\ytp{ y_2^{\prime} } 
\def\yttp{ \tilde y_2^{\prime}} 
\def\ythp{ y_3^{\prime} } 
\def\ythtp{ \tilde y_3^{\prime} }
\def\ptl{\partial}   

\begin{document}

\begin{flushright}
hep-th/0303173\\
BROWN-HET-1340
\end{flushright}

\vskip.5in

\begin{center}

{\LARGE\bf Massive IIA String Theory  and Matrix Theory Compactification }
\vskip 1in
\centerline{\large David A. Lowe, Horatiu Nastase and Sanjaye Ramgoolam}
\vskip .5in

\end{center}
\centerline{ Brown University}
\centerline{Providence, RI, 02912, USA}

\vskip 1in

\begin{abstract}

{ We propose a Matrix Theory approach to 
Romans' massive Type IIA supergravity. It is obtained 
by applying the procedure of Matrix Theory compactifications 
to Hull's proposal  of the Massive Type IIA 
String Theory as M-Theory on a twisted  torus. 
The resulting Matrix Theory is a super-Yang Mills theory 
on large $N$ three-branes with a space dependent non-commutativity 
parameter, which is also independently derived by a T-duality approach. We give evidence showing that 
the energies of a class of physical excitations 
of the super-Yang Mills theory show the correct 
symmetry expected from Massive Type IIA string theory 
in a lightcone quantization. 
}

\end{abstract}

\newpage

\section{Introduction}

Sometime ago Romans \cite{romans} found a massive deformation of
ten-dimensional Type IIA supergravity. This ten-dimensional theory
has remained something of a mystery from the string theory viewpoint.
Polchinski \cite{polchinski} argued this supergravity theory should lift to a massive Type IIA
string theory, corresponding to ordinary Type IIA string theory in the
background of a constant 10-form Ramond-Ramond field strength.

The problem of lifting this theory into M-theory has been considered by
a number of authors, including Hull \cite{hull}. His proposal is
similar in spirit to the idea of \cite{bs,ss} of obtaining
ten-dimensional Type IIB by compactifying M-theory on a 2-torus of
vanishing size but fixed complex structure. Instead one considers
M-theory compactified on a 2-torus bundle over $S^1$, $B(A,R)$, and
takes the limit of zero size. We review this construction in detail in
section 2.

We propose a nonperturbative formulation of M-theory in this
background using Matrix theory techniques. In section 3 we generalize
the construction of Seiberg and Sen \cite{seiberg,sen}, 
which provides us with a formulation of the discretized
light-cone quantization (DLCQ) of M-theory on the twisted torus.
The end result is a decoupled system of D3-branes in a background
$B_{\mu\nu}$ field, which we represent as a noncommutative Yang-Mills
theory with 8 linearly realized supercharges. An important novel
feature of our construction is the space dependent noncommutativity
parameter $\theta$.

In section 4, we construct the noncommutative Yang-Mills degrees of
freedom directly by compactifying an infinite system of D0-branes on
the twisted torus, following the general Matrix theory
\cite{bfss,halpern} procedure of
\cite{bfss,wat,grt,cds,ho}. Since we do not have 
the full zero-brane action in the original curved background 
we cannot proceed to derive the super-Yang-Mills theory as 
in the commutative case or the case with constant $\theta$. 
Nevertheless we obtain some useful information about the 
nature of fields in the theory. Concretely we give a 
construction of the covariant derivatives acting on an appropriate 
space of fields, and obeying the compactification constraints 
of the twisted torus. 

In section 5, we take advantage of known results 
about the star products in the presence of space-dependent 
non-commutativity and the result of section 3 
concerning the emergence of space dependent non-commutative Yang-Mills
in a generalized Sen-Seiberg limit in order to elaborate on the 
form of the action.  

The spectrum of states for a D8-brane background of massive Type IIA 
is examined in section 6 and we provide
evidence for an $SO(7)$ invariant spectrum of states, as
expected for DLCQ string theory in this background. This provides
further evidence supporting the Matrix formulation of the DLCQ string
theory.
In section 7 we consider a holographic dual spacetime to the
noncommutative gauge theory, generalizing \cite{hi,mr}, and we end with
conclusions and discussion in section 8. We comment 
on the extension of these Matrix compactification methods 
to other massive reductions of M Theory which admit de Sitter space 
solutions.

\section{Review of Hull's duality}

It is unknown how to lift the massive Type IIA string theory
\cite{hull} and its 
D8 background 
solution directly into M theory. While  M theory does not
seem to admit a cosmological constant,  a direct lifting of Romans'
massive ten-dimensional supergravity \cite{romans} would  yield
 an eleven-dimensional cosmological constant. 
One possible way around this is to obtain 
 the ten-dimensional mass  via a generalized
Scherk-Schwarz  reduction on a circle. 
The standard implementation of such a reduction requires 
a global symmetry in eleven dimensions. 
The action of the eleven dimensional supergravity 
 does not have such a symmetry but the equations of 
 motion do have a scaling symmetry,
which was exploited in \cite{lalupo} to reduce to a massive
ten-dimensional supergravity.
However, one obtains not Romans' massive supergravity but  a different
supergravity in ten dimensions.
 That massive supergravity can also be obtained as a 
usual reduction of a modified M theory \cite{hlw}.

Hull \cite{hull} was able to embed the massive supergravity 
\cite{romans} and the D8 background in M theory by introducing 
two extra T dualities, one of which is a
``massive T duality'' as defined in \cite{brgpt}. 
Let us describe this  in detail.
Scherk-Schwarz reduction is a mechanism for generating masses by
compactification in the presence of a global symmetry $\phi \rightarrow 
g(\phi)$, by an ansatz
\be
\phi (x^{\mu}, y) =g_y(\phi(x^{\mu}))~.
\ee
For the simplest case, of a $U(1)$ invariance, we can write
$\phi (x,y)=e^{2\pi i qmy}\phi(x)$, and obtain a mass $qm$ for
$\phi(x)$. 

In \cite{brgpt} it was shown that the Scherk-Schwarz reduction of 10d
IIB supergravity, using a $U(1)$ subgroup of the $SL(2,\mathbb{R})$ global
invariance is T dual to massive IIA supergravity, using a modified set 
of ``massive T duality'' rules. The reduction is given by
\be
g_y=
\begin{pmatrix} 1 & \frac{my}{R}\\ 0& 1
\end{pmatrix}
\ee
which implies
\be
\tau (x,y)\equiv a +i e^{-\phi} =\tau(x) +\frac{my}{R}~.
\ee
The monodromy (obtained for $y=R$) must be a symmetry of the full
quantum theory, that is it must be an element of $SL(2, \mathbb{Z})$, which 
implies that $m$ must be an integer. Then this compactification is
mapped by massive T duality into the usual compactification of the
massive IIA supergravity in ten dimensions. 

On the other hand, a Type IIB compactification on $S^1$ with nontrivial $\tau
(x,y)$ is equivalent to M theory compactified on a space $B$ which is a 
$T^2$ bundle over $S^1$, where the $T^2$ has modulus $\tau (x,y)$
fixed and area $A\rightarrow 0$. Equivalently, it is an F theory
compactification on $B$ where $A$ is fixed. 

We consider ten-dimensional massive Type IIA string theory, so the
 T dual (Type IIA) radius must go to 
infinity, hence the IIB radius $R$  goes to zero. 
If we also impose that $\tau (x) =\tau _0 =iR_2/R_1$, then massive IIA
supergravity is equivalent to M theory on the space $B(A,R)$, in the
limit $A\rightarrow 0, R\rightarrow 0$. The metric is (renaming $y$ as
$x_3$ and $R$ as $R_3$)
\be
ds_B^2= R_3^2 (dx_3)^2 +\frac{A}{{\rm Im} (\tau)}|dx_1 +\tau(x_3)dx_2|^2
=R_3^2 (dx_3)^2 + R_2^2(dx_2)^2 + R_1^2(dx_1+ m x_3 dx_2)^2
\ee
with all the radii going to zero, and the $x_i$ with periodicity 1, $x_i
\sim x_i+1$.
 In the limit, we should keep the massive IIA quantities 
fixed, so 
\be
g_s^A=\frac{l_s}{{\rm Im}(\tau_0)R_3}=\frac{R_1 l_s}{R_2 R_3}= {\rm fixed},
\;\; l_s =\frac{l_P^{3/2}}{R_1^{1/2}}={\rm fixed},\;\;
m \;\; {\rm fixed.}
\ee

A comment is in order regarding the quantization of the 10d IIA mass $m$
and the massive T duality. 
The relevant terms in the string frame supergravity actions are, 
for IIA 
\be
S_{IIA}=\frac{1}{k_{10}^2} \int \sqrt{g}(e^{-2\phi}R+\tilde{M}^2)
+...=\frac{1}{k^2}\int \sqrt{g} (e^{-2(\phi-\phi_0)}R
+(g_s^A)^2\tilde{M}^2)+...
\ee
whereas on the IIB side we have, similarly 
\be
S_{IIB}=\frac{1}{k^2}\int \sqrt{g} (e^{-2(\phi-\phi_0)}R+(g_s^B)^2
(\partial_{\mu} a)^2)+...
\ee
$\tilde{M}$ the supergravity mass parameter, is quantized in units of $1/l_s$, and it remains so when 
we reduce to 9d, whereas on the IIB side, $a=a_0+\frac{mx_3}{R_3}$, so 
the string frame masses are indeed equal
\be
 m_{(9)}^A\equiv M=g_s^A\tilde{M}=\frac{g_s^A m}{l_s}= \frac{m
  g_s^B}{R_3}= m_{(9)}^B
\ee

When talking about a duality, we have to specify the background as
well. The question is nontrivial, as the massive supergravity does not
admit a Minkowski background, not even a maximally supersymmetric
one. It does admit a half supersymmetric background, namely the D8
brane solution. 

The D8 has the string metric and dilaton ($d \sigma_{8,1}^2$ is the
$8+1$-dimensional Minkowski metric)
\bea
ds^2&=& H^{-1/2} (d\sigma_{8,1}^2) + H^{1/2} dx^2 \nonumber\\
e^{\phi}&=& H^{-5/4} \nonumber\\
H&=& c+|\tilde{M}| |x|=c+\frac{m}{l_s}|x|
\eea
where $c$ is an arbitrary constant of integration or (by the usual 
rescaling for p-branes)
\bea
ds^2&=& \bar{H}^{-1/2}(d\bar{\sigma}_{8,1}^2)+\bar{H}^{1/2} d\bar{x}^2
\nonumber\\
e^{\phi}&=& e^{\phi_0}\bar{H}^{-5/4}=g_s \bar{H}^{-5/4}
\nonumber\\
\bar{H}&=&H/c= 1+ g_s |\tilde{M}| |\bar{x}| =1+ \frac{g_s m}{l_s}|\bar{x}|
\eea
where $g_s$ is defined as the coupling constant at the position of the 
D8 brane.
The solution is obtained by promoting $\tilde{M}$ to a field
$\tilde{M}(x)$ and dualizing it to a 10-form field strength 
$F_{(10)}=\tilde{M}
\epsilon_{i_1...i_{10}} dx^{i_1}\wedge...\wedge dx^{i_{10}}$. Then 
$\tilde{M}(x)=\pm H'$, so the mass is piecewise constant, and jumps at
the positions of the D8 branes. The $\pm$ in the mass corresponds to
D8 branes vs. anti-D8 branes (since $F_{10}=*\tilde{M}$ is the field
strength for D8's), so for a D8 the supergravity mass jumps by a
positive amount, whereas for an anti-D8 by a negative amount. Note 
though that the tension of both is positive (the metric is the same
for both). 

On a compact space we should 
 think of  the D8's as being part of a D8-O8 system, 
with 16 D8's canceling the charge of the orientifold 8-plane O8. 
 If the transverse space is noncompact we can
assume that the O8 and the rest of the D8's
are far away, and concentrate on the local physics of 
 single, or coincident D8's.

We now have to find the M theory dual of the D8
solution. Dimensionally reducing to eight dimensions, one finds a 6-brane solution
(domain wall), which can be oxidized on the space $B(A,R)$ to the
Ricci-flat M
theory background 
\be
ds^2= H^{1/2}(H^{-1/2} d \sigma^2_{6,1} +H^{1/2} dx^2) +ds_B^2
=d \sigma^2_{6,1} +H dx^2 +ds_B^2
\label{mtheory}
\ee
where the moduli parameters of $ds^2_B$ are 
\be
R_3= H^{1/2}, \tau =m x_3 +iH~.
\ee
Equivalently, introducing the constants $r_i$, we define
\be
R_1= r_1/\sqrt{H}, ~R_2= r_2 \sqrt{H},~ R_3=r_3 \sqrt{H},
\label{radii}
\ee
and the limit becomes $r_i\to 0$ with
\be
\frac{r_1 l_s}{r_2r_3}= {\rm fixed, ~and} ~l_s~{\rm fixed.}
\ee
Counting parameters, we find 5 parameters in the M-theory
compactification, $l_P$, $R_i$ and $m$. This limit sends 2 parameters
to zero (e.g. $A=R_1 R_2\to 0$ then $R_3\to 0$), 
so that we are left with the 3 parameters of massive IIA,
$l_s$, $g_s$ and $m$.

\section{Matrix theory description in D8 background and T duality approach}

Hull's prescription tells us how to relate massive IIA string
theory to M theory. In this section we construct a Matrix description
of the M-theory compactified on $B(A,R)$.

The problem is nontrivial for two reasons. The first is that the space 
is curved, and moreover, if we write
\be
ds^2= (dz_3)^2+(dz_1+\alpha z_3 dz_2)^2 +(dz_2)^2
\label{metric}
\ee
we find the Ricci tensor components
\footnote{We have
 $g^{11}=1+\alpha^2 (z_3)^2, g^{22}=1$ and $g^{12}=-\alpha z^3$, and 
\bea 
&&\Gamma_{12}^3=-\frac{\alpha}{2} \;\;
 \Gamma_{13}^1=-\frac{\alpha^2z_3}{2}\;\;
\Gamma_{23}^1=\frac{\alpha}{2}(1-\alpha^2(z_3)^2)\nonumber\\
&&\Gamma_{22}^3 =-\alpha^2 z_3\;\; \Gamma_{13}^2= \frac{\alpha}{2} \;\;
\Gamma_{23}^2=\frac{\alpha^2 z_3}{2}
\eea
and all the rest are zero. Then we use
\be
R_{ab}=\partial_c \Gamma^c_{ab} -\frac{1}{2} \partial_a\partial _b ln
g + \frac{1}{2}  \Gamma^c_{ab}\partial_c ln g -\Gamma^c_{ad}\Gamma^d_{cb}
\ee
and the fact that $g=1$. }
\bea
R_{11}  =\frac{\alpha ^2}{2} &&
R_{12} = \frac{\alpha^3 z_3}{2} \nonumber\\
R_{22} = -\frac{\alpha^2}{2}(1
-\alpha^2(z_3)^2) && R_{33} = -\frac{\alpha^2}{2} ~.
\eea
The curvature scalar is 
\be
R=-\frac{\alpha^2}{2}~.
\ee
The metric (\ref{metric}) is invariant under the following isometries
\bea
T_1:&& z_1\rightarrow z_1+a_1, z_2\rightarrow z_2, z_3\rightarrow
z_3\nonumber\\
T_2:&&z_2\rightarrow z_2+a_2, z_1\rightarrow z_1, z_3\rightarrow
z_3\nonumber\\
T_3:&&z_3\rightarrow z_3+a_3, z_1\rightarrow z_1-\alpha z_2a_3, 
z_2\rightarrow z_2
\eea
with Killing vectors $V_1=\partial _1$, $V_2=\partial_2$ and $V_3=
\partial_3-\alpha z_2\partial_1$. We also note that $[T_2,T_3]\neq 0$.
By identifying under the isometries with $a_i=R_i$ we obtain the space
$B(A,R)$, and then we have (since $z_i=R_i x_i$)
\be
\alpha= \frac{mR_1}{R_2 R_3} =M~.
\ee
We note therefore that we can trust supergravity as long as 
\be
\alpha l_P=\frac{mR_1l_P}{R_2R_3} \ll 1
\ee
which is true in our limit ($\frac{R_1}{R_2R_3}$ fixed,
$l_P\rightarrow 0$).


In the following we choose to work with the D8 background,
corresponding to the M theory metric (\ref{mtheory}) with radii
(\ref{radii}). We propose a Matrix description is obtained by
considering the action of $N$ D0-branes in the D8 background
(\ref{mtheory}). Since the radii of $B$ go to zero, we have to make T dualities
in the 3 directions of $B$, and so the Matrix model describing massive
IIA will be the action of $N$ D3 branes in the T dual background.
It is understood that the general procedure used 
to obtain the dual Matrix model will be the same for any massive IIA 
background with a light-like symmetry.

First, however, we must define correctly the limit taken on the M
theory, and see what kind of limit we obtain for the D3 brane. This 
is described in detail in the appendix, but we will give here only the 
relevant facts. 
Sen \cite{sen} and Seiberg \cite{seiberg} give a prescription for 
the discretized light-cone quantization (DLCQ) of M theory (with light-like radius R and finite $l_P$) 
on a torus of finite radii $R_i$. One goes to an equivalent $\bar{M}$ 
theory with $\bar{l}_P\rightarrow 0$ and spacelike 11th direction of 
radius $R_s\rightarrow 0$ and compactification radii
$\bar{R}_i\rightarrow 0$ such that 
\be
\frac{R_s}{\bar{l}_P^2}=\frac{R}{l_P^2},\;\;\;
\frac{\bar{R}_i}{\bar{l}_P} =\frac{R_i}{l_P}
\ee
are held fixed in the $\bar{l}_P\rightarrow 0$ limit. Then one makes T
dualities in the compact directions and gets a decoupled theory of 
Dp-branes (D3 in our case) with finite $g_{YM}$ and dual radii
\bea
\tilde{R}_i &=& \frac{\bar{l}_s^2}{\bar{R}_i}=\frac{l_P^3}{R_iR}
\nonumber\\
\tilde{g}^2_{YM}&=& \tilde{g}_s=\frac{R^3}{l_P^6}\prod_i\tilde{R}_i~.
\eea

As we can see, the parameters of the dual D3 matrix model do not depend on 
the parameters of the $\bar{M}$ theory, which was introduced just to 
prove the duality. Therefore we can apply another limit to this
construction (independent of the Sen-Seiberg $\bar{l}_P\rightarrow 0$
limit), namely $l_P\rightarrow 0$ and $R_i\rightarrow 0$, with
$l_P^3/R_1=(l_s^A)^2=l_s^2$ and $R_1/(R_2R_3)$ kept fixed. We also need to 
make a ``9-11 flip'', namely to reinterpret the lightcone coordinate $R$
as the 11th direction (since in the M theory construction of massive
IIA $R_1$ takes the role of 11th coordinate).

The parameters of the \SYM\ are
\be
\tilde{R}_1 =\frac{l_s^2}{R}, ~\tilde{R}_2 =\frac{R_1}{R_2}\tilde{R}_1,~
\tilde{R}_3=\frac{R_1}{R_3}\tilde{R}_1, ~\tilde{g}_s=\tilde{g}_{YM}^2
=\frac{l_sg_s^A}{R_1}
\ee
and the inverse relations are, if $R=Nl_s$,
\bea
&& R=Nl_s~, \;\;\; R_1 = \frac{N}{g_{YM}}\sqrt{\tilde{R}_2\tilde{R}_3} 
\nonumber\\ && g_s^A
=\sqrt{\tilde{g}_{YM}^2\frac{\tilde{R}_2\tilde{R}_3}{\tilde{R}_1^2}
},\;\;\; l_s = N \tilde{R}_1~.
\eea
In order to still have decoupling of the string theory from the D3
brane theory, we need to have $\bar l_s\to 0$ and the S-dual string length $\tilde{g}_s{\bar{l}}^2_s\rightarrow 0$, 
which is satisfied in the $\bar{M}$ theory, since
\be
\bar{l}_s^2=\frac{l_P^3}{R}\frac{\bar{l}_P}{l_P} \Rightarrow 
\tilde{g}_s\bar{l}_s^2=\frac{l_s^3 g_s^A}{R}\frac{\bar{l}_P}{l_P}\to 0~.
\ee

Let us now follow this procedure in order to find the Matrix model 
description of the background (\ref{mtheory}): 9-11 flip, going to 
$\bar{M}$ theory, dimensional reduction to string theory, followed by 3
T dualities. The string theory background in the $\bar{M}$ theory is
\be
ds^2= d \sigma^2_{5,1} +H dx^2 +ds_B^2
\label{twoa}
\ee
with the radii given in (\ref{radii}) and constant dilaton $\phi_0$, 
and now we need to perform 3 T dualities. We will concentrate on the 
space $B$, with metric
\be
ds_B^2= H (\bar r_3^2dx_3^2 +\bar r_2^2dx_2^2) +\frac{\bar r_1^2}{H}(dx_1 + mx_3
dx_2)^2
\ee
and work  with string metrics, on which 
the T dualities act in a simple way. We will work in units of
$\bar l_s$. If we want to restore the $\bar l_s$ 
dependence we can formally put $\bar r_i \rightarrow \bar r_i/ \bar l_s, x_i
\rightarrow x_i \bar l_s, m \rightarrow m/\bar l_s$. 

The Buscher T duality rules \cite{buscher1,buscher2,buscher3} are
\bea
\hat{g}_{00}&=&\frac{1}{g_{00}}\nonumber\\
\hat{g}_{0i}&=& \frac{B_{0i}}{g_{00}}\nonumber\\
\hat{g}_{ij}&=& g_{ij}-\frac{g_{0i}g_{0j}-B_{0i}B_{0j}}{g_{00}}
  \nonumber\\
\hat{B}_{0i}& =& \frac{g_{0i}}{g_{00}}\nonumber\\
\hat{B}_{ij}& =& B_{ij}+ \frac{g_{0i}B_{0j}-B_{0i}g_{0j}}{g_{00}}
\nonumber\\
\hat{\phi}&=& \phi-\frac{1}{2} \log (g_{00})
\eea
Here the coordinate 0 of the T duality is defined such that
$\partial_0$ is the Killing vector of an isometry.
It is worth noting here that one might be worried that we have to use
the massive T duality rules at some point, however the $m$-dependent
terms are only in the transformation rules of the RR fields (see 
\cite{brgpt}). 

As  we saw, we have 3 isometries, $T_1, T_2, T_3$.  $T_2$ and 
$T_3$ do not commute, so the  order of  T
dualities matters. We will choose to do $T_1$, then $T_2$, then $T_3$. 
We begin by considering the simpler case of T dualities on the twisted
torus with 
the radii and $\bar l_s$, 
$H$ set to $1$ corresponding to the core of the D8 background $x=0$,
and later we will generalize this to the complete background. 
After $T_1$ we have  : 
\bea
ds^2&=& (  dx_3^2 +  dx_2^2 
+ dx_1^2)\nonumber\\
B_{12}&=& m x_3 \Rightarrow H_{123}=m \nonumber\\
e^{\phi}&=& e^{\phi_0}~.
\label{aftert1}
\eea
After $T_2$ we have 
\bea
ds^2 &=& ( dx_3^2 + dx_1^2) + (dx_2 + m
x_3 dx_1)^2 \nonumber\\
e^{\phi}&=& e^{\phi_0} ~.
\label{aftert2} 
\eea
$T_3$ is generated by the vector 
$V_3=\partial_3 -m x_1\partial_2$. Transforming to coordinates 
\be
x_3'=x_3, x_2'=x_2 + m x_1x_3
\label{coordtr1}
\ee
implies $V_3=\partial'_3$
so that we can apply the usual T duality rules.

The metric in the new coordinates (after dropping primes on 
coordinates) 
\bea
ds^2 &=& ( dx_3^2 + dx_1^2) + (dx_2  - m
x_1 dx_3  )^2 \nonumber\\
e^{\phi}&=& e^{\phi_0} ~.
\label{aftercoord} 
\eea
After the third $T$-duality, we have 
\bea
ds^2 &=& dx_1^2  +  \frac{(dx_2^2
  + dx_3^2  )}{ 1 + m^2 x_1^2 }    \nonumber\\
B_{23}dx^2 \wedge dx^3&=& - \frac{m x_1}{1+ m^2 x_1^2 } dx_2 \wedge dx_3
 \nonumber\\
e^{\phi} &=& \frac{e^{\phi_0}}{ (1+ m^2 x_1^2 )^{1/2}}~.
\label{aftrt3}
\eea
The open string metric and $\theta$-field 
are 
\bea 
ds^2 &=& dx_1^2 + dx_2^2 +dx_3^2 \nonumber\\
\theta^{23} &=& m x^1 ~.
\label{opfin} 
\eea

The closed string metric in (\ref{aftrt3}) is no longer periodic 
in $x_1$. The metric in (\ref{aftert2})  has the property 
that the $(23)$ torus at $x_1+1$ is related to 
that at $x_1$ by an $SL(2,\mathbb{Z})$ transformation 
\bea 
 A  = \begin{pmatrix} 1&0 \\ -m&1  \end{pmatrix}~.
\eea
This $ 2 \times 2  $ matrix is embedded 
in the full $O(2,2; \mathbb{Z} )$ T-duality 
group as 
\bea 
 S  = \begin{pmatrix} A & 0 \\  0 & {A^{T}}^{-1} \end{pmatrix}~.
\eea
as explained for example in the review  \cite{gipora}.  

The closed string metric and B-field 
in (\ref{aftrt3}) obey the property that the 
background matrix $ E = G + B $ of the 
$(23)$ torus and the dilaton are related 
\bea
E ( x_1 +1 ) = \frac{  a E(x_1) + b }{  cE(x_1) + d } \nonumber\\
e^{\phi( x_1 + 1 )}  = e^{\phi ( x_1 ) } 
\left( \frac {{\rm det} ~g(x_1 + 1 ) }{{\rm  det} ~g( x_1)} \right)^{1/4}  
\label{O22trans}  
\eea 
where $a,b,c,d $ are $2 \times 2 $ 
matrices entering a $4 \times 4 $ 
$O(2,2) $ matrix 
\bea
M &=& \begin{pmatrix} a&b \\  c&d \end{pmatrix} ~.
\eea 
One easily checks (\ref{O22trans}) when $x_1 =0$ 
and with a little more work for general $x_1$. 
The $ a,b,c,d $ are calculated by 
observing that the 
shift by $x_1$ in (\ref{aftrt3})  can be accomplished by 
first T-dualizing to (\ref{aftercoord}), doing the shift 
and T-dualizing back. 
The $O(2,2; \mathbb{Z} )$ matrix $T_3$ for the T-duality
along $x_3$ in (\ref{aftercoord}) is 
\bea
T_3 &=& \begin{pmatrix}1 & 0 & 0 & 0 \\
                      0 & 0 & 0 & 1 \\
                      0 & 0 & 1 & 0 \\ 
                      0 & 1 & 0 & 0   \end{pmatrix}
\eea 
and $ M = T S T^{-1} $. This gives 
\bea
a &= & \begin{pmatrix} 1&0 \\ 0 &1 \end{pmatrix} \nonumber \\
b &= & \begin{pmatrix}0&0 \\ 0 &0 \end{pmatrix}   \nonumber \\
c &= & \begin{pmatrix}0&m \\ -m &0\end{pmatrix}\nonumber \\
d &= & \begin{pmatrix}1&0 \\ 0 &1 \end{pmatrix}~.\nonumber  \\
\label{abcd}
\eea 
It is also interesting to observe that the 
open string background 
in (\ref{opfin}) characterized by the matrices $G$ and $\Theta$  transforms
under a shift of $x_1$ by the same $O(2,2; \mathbb{Z} )$ 
matrix $M$ when we use the action of $O(2,2; \mathbb{Z} ) $
given by Seiberg-Witten : 
\bea 
G(x_1 + 1 ) &=& G(x_1) = \left( a + b \Theta(x_1)  \right) G( x_1)  
\left( a + b \Theta(x_1)  \right)^T   \nonumber  \\
\Theta ( x_1 + 1 )   &=&  ( c + d \Theta (x_1) ) ( a + b \Theta (x_1))^{-1} 
\nonumber \\
g_{YM} (x_1 +1) &=&  
 =g_{YM} (x_1) \left( {\rm det} ( a +  b \Theta (x_1)) \right)^{\frac{1}{4} } 
 \nonumber  \\ 
\eea
The final background is to be viewed as 
a $T^2$ bundle over $S^1$ where the 
$T^2$ is twisted by an element of the full 
$T$-duality group of the torus upon transport along the $S^1$. 
This structure of  the closed string background obtained 
after T-dualizing the twisted torus has been observed recently 
in \cite{kstt} and related work appears 
in \cite{dh,hmw}.

We now describe the T-dualities on the full background. 
The above remarks on the $O(2,2)$ carry over. 
After T duality on $T_1$ we have (the full 10d metric) 
\bea
ds^2&=&  d \sigma^2_{5,1} +H(dx^2 +\bar r_3^2 dx_3^2 +\bar r_2^2 dx_2^2 
+1/\bar r_1^2 dx_1^2)\nonumber\\
B_{12}&=& m x_3 \Rightarrow H_{123}=m \nonumber\\
e^{\phi}&=& \frac{e^{\phi_0}}{\bar r_1} \sqrt{H}
\label{nsfive}
\eea
which we recognize as nothing but the NS5 brane metric smeared over 
the transverse directions 1,2,3. 

After T duality on $T_2$ we have 
\bea
ds^2 &=& H(\bar r_3^2dx_3^2 +1/\bar r_1^2 dx_1^2) +\frac{1}{\bar r_2^2 H}(dx_2 + m
x_3 dx_1)^2 \nonumber\\
e^{\phi}&=& \frac{e^{\phi_0}}{\bar r_1\bar r_2}
\eea
which is the same metric as we started from, with inverted radii $\bar r_1,
\bar r_2$ and with 1 and 2 interchanged. 

Then in these dual coordinates, $T_3$ has Killing vector
$V_3=\partial_3 -m x_1\partial_2$. Applying the coordinate
transformation (\ref{coordtr1}) 
 and $T_3$ T duality we get
(restoring also the $\bar l_s$ dependence for later use) 
\bea
ds^2 &=& \bar l_s^4 \left(H (dx_1^2/\bar r_1^2) + \frac{H^{-1} (dx_2^2/\bar r_2^2
  +dx_3^2/\bar r_3^2)}{ 1+  \left( \frac{m \bar l_s^2  x_1}{H \bar r_2 \bar r_3}\right)^2}\right)\nonumber\\
B_{23}dx^2 \wedge dx^3&=& -\bar l_s^4\frac{m \bar r_1}{\bar r_2\bar r_3H^2}\frac{x_1/\bar r_1}{
1+ \left( \frac{m \bar l_s^2  x_1}{H \bar r_2 \bar r_3}\right)^2 } dx_2/\bar r_2 \wedge dx_3
/\bar r_3 \nonumber\\
e^{\phi} &=& \frac{ \bar l_s^3e^{\phi_0}} {\bar r_1\bar r_2\bar r_3} H^{-1/2} 
\left(1+\left( \frac{m \bar l_s^2  x_1}{H \bar r_2 \bar r_3}\right)^2
\right)^{-1/2}~.
\label{closed}
\eea
Let us now define $\tilde{y}_i=\bar l_s^2x_i/\bar r_i$ and calculate open 
string variables, to find the metric $G$ and noncommutativity
parameter $\theta$ the D3-brane sees \cite{sw}. 
Using
\be
\left(G+\frac{\theta}{\bar l_s^2}\right)^{ij}=\left(\frac{1}{g+\bar l_s^2B}\right)^{ij}
\ee
we get  for the full 10d metric
\bea
ds^2 &=& d\sigma^2_{5,1} + H(dx^2 + d\tilde{y}_1^2) +\frac{d
\tilde{y}_2^2 +d\tilde{y}_3^2}{H} \nonumber\\
e^{\tilde\phi}&=& \frac{\bar l_s^3e^{\phi_0}}{\bar r_1\bar r_2\bar r_3 \sqrt{H}}\nonumber\\
\left[ \tilde{y}_2, \tilde{y}_3 \right] &=&-i (m
\frac{\bar r_1}{\bar r_2\bar r_3})\bar l_s^2
\tilde{y}_1= -i\tilde{\alpha}\tilde{y}_1\nonumber\\
H&=&1+ \frac{m \bar r_1}{\bar r_2 \bar r_3}|x|= 1+ \frac{m \bar r_1 \bar l_s^2}{\bar r_2 \bar r_3} |X|
=1+\tilde{\alpha} |X|~,
\label{noncomm}
\eea
where we have defined $X= x/\bar l_s^2$.
 
Recalling that 
$\bar{l}_s$ goes to zero in the infinite boost limit, and then, with 
$\tilde{y}_i$ and $\tilde{\alpha}$ fixed, we have  
\be
{\bar l_s}^2\sim \epsilon^{1/2}, \;\;\; g_{ij} \propto 
{\bar l_s}^4\sim \epsilon,
\;\;\; G_{ij} \sim \frac{g_{ij}}{ {\bar l_s}^4}={\rm fixed} 
\ee 
which is nothing other than the Seiberg-Witten limit for
noncommutative geometry, which means that the theory on the D3 branes 
is nothing other than noncommutative super Yang-Mills theory with
variables (metric, dilaton and noncommutativity) given in
(\ref{noncomm}). 
The noncommutativity 
parameter is
\be
\tilde{\alpha}=\frac{m \bar{r}_1 \bar{l}_s^2}{\bar{r}_2\bar{r}_3}
=\frac{m r_1}{r_2r_3}\frac{l_P^3}{R}=
\frac{mg_s^A}{l_s^A}\frac{l_P^3}{R} =\frac{m\tilde{R}_2\tilde{R}_3
}{\tilde{R}_1}~.
\label{ncparam}
\ee

\section{ Space-dependent non-commutativity and solution of 
 Matrix Theory constraints  }

Let us now see that the \SYM\ defined on (\ref{noncomm}) can also be 
obtained from an algebraic approach in Matrix theory. Our goal will be
to reproduce the noncommutative structure of the Matrix degrees of
freedom $X^i(t)$. We follow the general procedure for 
compactification \cite{bfss,wat,grt,cds,ho}. We need to find T-dual 
variables $y_i$ such that the matrices $X^i$ can be represented as
covariant derivatives. For the simple case of circle compactification, 
one represents the algebra
\be
\Omega X \Omega ^{-1}=X +R \label{ident}
\ee
by $X=iD_y \equiv i\partial_y +A , \Omega=e^{iRy}$, since $[i\partial_y +A,
e^{iRy}]= - R e^{iRy}$. 

Let us review this in a bit more detail. We start 
with a finite number $k$ of zero branes on a circle. 
This is equivalent to having an infinite number of copies 
of $k$ zero branes along a line, with zero branes 
separated  by a constant  periodic shift  $R$ related by 
a gauge transformation $ \Omega $ in $ U(\infty)$ as in
(\ref{ident}). 
One writes  $X^{ab}$ matrices as  $X^{a'b'}_{mn}$ 
where $a^{\prime}, b^{\prime} $ run over the $k$ zero branes
and $m,n$ are integers. 
These matrices are now operators on states labeled by 
an integer $m$ 
corresponding to $e^{im Ry} = |m\rangle  $, tensored by finite 
$k \times k $ matrices. The operators on the states $|m> $ 
can be viewed as the algebra of functions on 
 the T dual space. The form  $A(y)=\sum_p A_p e^{ipRy}$, 
allows us to read off the T-dual radii. The matrix elements 
of $X$ and $\Omega $ are given by 
  $X_{n,m}=-nR\delta_{n,m}+A_p\delta_{n,
  m+p}$ and  $\Omega_{nm}=\delta_{n,m+1} $. 
So we have the T dual Matrix model description in terms of D1-branes
(and by generalization, Dp-branes). Fluctuations in the compact $X$ are
mapped to fluctuations in the gauge field $A$, and part of the
original 
Matrix degrees of freedom were 
 used to generate functions of  the worldvolume direction $y$. 


We will now try and apply this procedure to our case. We will treat
first the case when the harmonic function $H=1$, and will see later what 
complications H introduces. We can think of it as working in the near core
region $x\simeq 0$. We will also put for simplicity $R_i=1$ for the 
moment, and return to the general case later on.  

In our case we have 3 isometries, $T_1, T_2, T_3$, which means that 
we need to impose constraints on the D0 Matrix model analog to
(\ref{ident}) and try to solve them in terms of a T dual space.
The constraints are
\bea
T_1:&& \Omega_1 X_1 \Omega_1^{-1}=X_1+1
\nonumber\\
&&\Omega_1 X_2 \Omega_1^{-1}=X_2\nonumber\\
&& \Omega_1 X_3 \Omega_1^{-1}=X_3
\label{xone}
\eea
where $\Omega_1$ is a transformation acting on the X's which
corresponds to the isometry $T_1$, and similarly
\bea
T_2:&& \Omega_2 X_1 \Omega_2^{-1}=X_1
\nonumber\\
&&\Omega_2 X_2 \Omega_2^{-1}=X_2+1\nonumber\\
&& \Omega_2 X_3 \Omega_2^{-1}=X_3 \nonumber\\
T_3:&&\Omega_3 X_1\Omega_3^{-1}= X_1-m X_2\label{dercom}\\
&&\Omega_3 X_2 \Omega_3^{-1}=X_2 \nonumber\\
&&\Omega_3 X_3 \Omega_3^{-1} =X_3 +1 ~.\label{xthree} 
\eea
We noted that $T_2$ and $T_3$ don't commute, and we can see that 
therefore $\Omega_2$ and $\Omega_3$ don't commute. Namely, 
if we put $M=\Omega_2^{-1} \Omega_3^{-1} \Omega _2 \Omega_3$, 
then we have 
\bea
M X_1 M^{-1} &=& X_1 -m \nonumber\\ 
M X_2 M^{-1} &=& X_2 \nonumber\\
M X_3 M^{-1} &=& X_3~.
\label{commut}
\eea

{}From the relations defining $T_1$ and $T_2$ we can see that we can put 
$X_1= iD_1, X_2= iD_2$ and $\Omega_1=e^{iy_1} , \Omega_2=e^{iy_2} $ just
as in flat space. The commutation of relation of $ \Omega_3 $ 
with $X_3$ can also be solved by $ X_3 = i D_3 $ and $ \Omega_3 = e^{i
  \alpha y_3} $ ( here $ \alpha = m $ ). 
The relations (\ref{commut}) allow us to  solve for $M=e^{-i m 
  y_1}$. Hence we deduce 
that $y_2$ and $y_3$ don't commute, and we get exactly the noncommutativity
relations 
\be
[y_2, y_3]= i \theta_{23} =  i\alpha y_1, \;\; [y_1, y_3]=[y_1, y_2]=0
\label{noncom}
\ee
as we obtained from the T duality approach of the last section.

The relations (\ref{noncom}) also imply nontrivial
commutations for derivatives and coordinates. Indeed, by
taking the commutator with various derivatives of the relations 
(\ref{noncom}), we get a set of equations for $[\partial_i, y_j]$. We
will not list them here but just mention a  solution, namely
$[\partial_i,y_i]=1$ as usual, but also $[\partial_1, y_3]=-i\alpha
\partial_2$. The relevant equation is obtained from $[\partial_1, [y_2,y_3]]$
\be
[[\partial_1, y_2], y_3]+[y_2, [\partial_1, y_3]]=-i\alpha
\label{noncomtwo}
\ee
and we can see that it is indeed solved by $[\partial_1, y_3]=-i\alpha
\partial_2$.

It is useful to observe that a 
change of variables maps the non-commutativity parameter 
to a constant. 
Indeed, in  the open string metric 
\be
ds^2= dy_1^2 +dy_2^2 +dy_3^2+ds_{tr}^2
\ee
with the noncommutativity $\theta_{23}= \alpha y_1$ we can make the
change of variables $y_1=y'_1, y_2=y'_2 y'_1, y_3=y'_3$, 
after which the theory has metric
\be
ds^2={dy'_1}^2 +{dy'_3}^2 +d(y'_1y'_2)^2
\ee
and constant noncommutativity ${\theta '}_{23}=\alpha$. The closed string
metric and B field  in (\ref{closed}) become, under the transformation:
\bea
ds^2&=&
 {dy'_1}^2(1+\frac{{y'_2}^2}{1+\alpha^2{y'_1}^2})+\frac{{dy'_3}^2+{y'_1}^2
{dy'_2}^2+2y'_2y'_1dy'_2dy'_1}
{1+\alpha^2{y'_1}^2}
\nonumber\\
l_s^4B&=& \frac{\alpha y'_1(y'_1dy'_2+y'_2dy'_1)\wedge dy'_3}{1+\alpha^2{y'_1}^2}~.
\eea
To put back the $R_i$ and $l_s$ dependence, we only need to substitute
$\alpha =\tilde{\alpha}/l_s^2$ in the above.

 Since the noncommutativity is constant in the new coordinates $y_i'$, 
 we can realize the  commutation relations
 \bea
 \left[ y'_i, y'_j \right] &=& i \theta'_{ij} \nonumber \\
 \left[ \partial'_i, y'_j \right] &=& \delta^{i}_{j}  \nonumber \\
 \label{crpp}
 \eea
as we explain further below. 
The commutation relations (\ref{crpp}) imply the
following relations for the unprimed coordinates
\bea
\left[ y_i, y_j \right] &=& i \theta_{ij}(y_1)  = 
 i \theta^{\prime}_{ij}(y_1)   \nonumber \\
\left[\partial_i, y_j \right] &=& \delta^i_j + \Psi^i_j(\partial)   
\nonumber\\
\label{crup}
\eea
where $\theta_{ij}(y) $ has nontrivial components $\theta_{23} =-\theta_{32}=
\alpha y_1$, $\Psi^i_j$ has nontrivial components $\Psi^1_3 = i
\alpha \frac{\partial}{\partial y_2}$, 
as we wanted (see (\ref{noncom}) and (\ref{noncomtwo}) ).
These guarantee 
that if we set $X_i = i \frac{\partial}{\partial y_i }$ 
and $ e^{ i y_i } $ we correctly obey the constraints
in (\ref{xone})(\ref{dercom})(\ref{xthree}). This provides the 
foundation for 
the  general solution including gauge fields but 
we first need to review the construction of the 
covariant derivatives including gauge fields in the case 
of constant non-commutativity.   



We recall some facts about the construction of 
a  non-commutative gauge theory from 
covariant derivatives in the context of an ordinary 
constant noncommutativity $\theta^{\prime}_{ij}$. To have notation 
which agrees with our set-up we will let 
$i,j$ run over $2,3 $ and we will let the non-commutative torus 
algebra be generated by $ \ytp, \ythp$. Consider compactification 
constraints generated by    
$ \Omega_2^{\prime}  = e^{i \ytp } $ and $ \Omega_3^{\prime}  = e^{i \ythp } $ 
where $ \Omega_i'  \Omega_j'  (\Omega_i')^{-1} (\Omega_j')^{-1} = 
e^{ i \theta^{\prime}_{ij} } $, 
with the only non-trivial components being $ \theta'_{23} = -
\theta'_{32} = \theta $. The $ \Omega_i^{\prime}$ and $ X_i^{\prime}$ are 
 represented in a Hilbert space where there is 
non-trivial commutant  generated by 
$ \tilde \Omega_2^{\prime} , \tilde \Omega_3^{\prime} $ (i.e. $[\Omega'_i,
\tilde \Omega'_j] =0$) which have a 
non-commutativity parameter $ \tilde \theta = - \theta$.
Writing $ \tilde \Omega_2' = e^ {i  \yttp } $ 
and $\tilde \Omega_3' = e^ { i \ythtp  } $, we have 
$ [ \yttp , \ythtp ] = - i \theta $. Explicit construction 
of  the $y_i^{\prime} $ and $ \tilde y_i^{\prime} $'s or 
equivalently the $\Omega_i^{\prime}$
and $ \tilde \Omega_i^{\prime} $ in terms of coordinates $w_2,w_3$ 
which commute with each other and satisfy standard commutation 
relations with their derivatives $ \frac{\ptl}{\ptl w_2} , 
\frac{\ptl}{\ptl w_3}$ and together describe  a four-dimensional
 phase space are given in  (\cite{sw})
\bea
\ytp =  w_2 +  \frac{i \theta}{2} \frac{\ptl}{\ptl w_3 }   \nonumber \\
\ythp = w_3 -  \frac{i \theta}{2} \frac{\ptl}{\ptl w_2}       \nonumber \\ 
\yttp =  w_2 -  \frac{i \theta}{2} \frac{\ptl}{\ptl w_3 }  \nonumber \\ 
\ythtp = w_3 +  \frac{i \theta}{2}  \frac{\ptl}{\ptl w_2 } ~.\nonumber \\
\label{phsprl} 
\eea
These formulas can be used to check that
the correct mutual non-commutativity of $y_i^{\prime} $ and of 
$\tilde y_i^{\prime}$ are reproduced, as well as the vanishing 
commutators of any $y_i^{\prime} $ with $ \tilde y_i^{\prime} $.  

We can define derivatives with respect to $y_i^{\prime} $ and $ \tilde
y_i^{\prime} $  
\be 
[ \frac{\partial}{\partial y_i^{\prime} }, y_j^{\prime} ]=\delta_i^j, \;\;\; 
[ \frac{\ptl}{\ptl y_i^{\prime}  }, \tilde y_j^{\prime} ]
=0
\ee
and 
\be 
[ \frac{\ptl}{\ptl \tilde y_i^{\prime} }, \tilde y_j^{\prime} ]=
\delta_i^j, \;\;\; 
 [ \frac{\ptl}{\ptl \tilde y_i^{\prime} }, y_j^{\prime} ]
=0~.
\ee
These constraints can be solved by defining the derivatives 
in terms of appropriate commutator actions with elements 
in the algebra of $y, \tilde y $ 
\bea
 &&\frac{\ptl}{\partial y_i^{\prime} }
=-i(\theta^{-1})^{ij} y_j^{\prime} , \nonumber\\
&&\frac{\ptl}{\ptl {\tilde{y}_i^{\prime}}}
 = +i (\theta^{-1})^{ij} \tilde{y}_j^{\prime} 
\eea
and the non-trivial commutation relations of derivatives follow
\bea
&& [ \frac{\ptl}{\partial y_i^{\prime} }, \frac{\ptl}{\ptl
   y_j^{\prime} } ]
 =-i (\theta^{-1})^{ij} \nonumber\\
 &&[ \frac{\ptl}{\ptl \tilde y_i^{\prime} }, \frac{\ptl}{\partial
     \tilde y_j^{\prime}  }  ] 
 = + i(\theta^{-1})^{ij} ~.
\eea 
The fact that the derivatives can be expressed in terms 
of commutator action with elements in the algebra plays 
an important role in  \cite{seiberg2} 
in the context of a discussion of solutions of Matrix 
Theory describing extended objects in $R^2$ and having 
a non-commutative worldvolume theory derived from Matrix Theory. 

The presence of the commutant generated by the $\tilde y^{\prime}  $ is 
important in getting solutions to the constraints with 
non-trivial gauge fields. The simplest gauge theories 
are in fact obtained when we take the covariant  
derivatives to be  $ \partial_{y_i^{\prime}  } + 
\partial_{\tilde y_i }  - i A_i ( e^{i \tilde y_i } )  $.
Such  a choice of derivative  was implicit
for example in \cite{ho}.  It is useful to note that 
\be 
 [ \frac{\ptl}{\partial y_i^{\prime}  } +  
\frac{\ptl}{\partial \tilde y_i^{\prime}  } , 
\frac{\partial}{\partial y_j^{\prime} } +  
\frac{\ptl}{\partial \tilde y_j^{\prime}  } ] = 0 
\ee
which means that there is no background magnetic field.
As far as solving the compactification constraints 
we could work with a more general set of partial 
derivatives  $ ( \partial_{y_i^{\prime}  } + \phi_{ij}
 \partial_{\tilde y_i^{\prime}  } ) $.


We can now write a solution for the $X$ operators
acting on a Hilbert space of functions. 
Since the periodicities are simple in the 
$y$-coordinates, we are lead to consider
functions of generated by $e^{i y_1 }$, $e^{i  y_2 }  $ and $e^{i   y_3 } $.
Recalling the discussion above, where we saw that 
the constraints are expressed in terms of $y^{\prime} $ variables 
whereas the fields are functions of variables 
$ \tilde y^{\prime} $, we are lead to work with  the Hilbert 
space of functions of the form 
\be   
\psi = \sum_{n_1, n_2, n_3} 
 \psi_{n_1, n_2, n_3 } e^{i  n_1 y_1^{\prime}  }   
e^{i  n_2 y_1^{\prime} \tilde y_2^{\prime  }}
  e^{i  n_3 \tilde y_3^{\prime}  }  
\ee 
where $ n_i$ are arbitrary integers. 
The $y_i $ and $ \tilde y_i^{\prime} $,
for $ i = 2,3 $ are constructed in 
terms of a four dimensional phase space as in 
(\ref{phsprl}). 

On this Hilbert space we can write operators 
\bea 
&& X_1 = i \frac{\partial}{\partial y_1^{\prime} } - i y_2^{\prime} 
\left(  \frac{1}{y_1^{\prime}  }  \frac{\partial}{\partial y_2^{\prime} }
  - i A_2 ( e^{iy_1^{\prime} } , e^{ i \yop \yttp} ,  
e^{ i \ythtp } )\right)  -  \frac{ i \yttp}{\yop} \frac{\partial}{\partial \yttp}
 + A_1  ( e^{iy_1^{\prime} } , e^{ i \yop \yttp} ,  e^{ i \ythtp } )
\nonumber\\
&& X_2 = \frac{i}{\yop} \frac{\ptl}{\ptl \ytp} + 
A_2 ( e^{iy_1^{\prime} } , e^{ i \yop \yttp} ,  e^{ i \ythtp } ) 
\nonumber\\
&& X_3 = i\frac{ \ptl}{\ptl \ythp} + 
A_3 ( e^{iy_1^{\prime} } , e^{ i \yop \yttp} ,  e^{ i \ythtp } )
\nonumber\\
&&\Phi^a = \Phi^a (  e^{iy_1^{\prime} } , e^{ i \yop \yttp} 
,  e^{ i \ythtp } )~.
\nonumber\\
\label{solconsts}  
\eea
The constraints are generated by
\bea  
 && \Omega_1 = e^{ i y_1^{\prime} } \nonumber\\  
 && \Omega_2 = e^{ i y_1^{\prime} y_2^{\prime}  } \nonumber \\   
 && \Omega_3 = e^{ i y_3^{\prime}  } ~.\nonumber \\
\label{omegasyp} 
\eea  
With these expressions we can check that the constraints 
in (\ref{xone}), (\ref{dercom}) , (\ref{xthree}) are satisfied and 
that the $X$'s correctly act in the Hilbert space. 
We elaborate on some aspects of these properties.  
The combination  $ ( i \frac{\partial}{\partial y_1^{\prime} } 
-  \frac{ i \yttp}{\yop} \frac{\partial}{\partial \yttp} ) $ is
necessary in $X_1$ because it allows $X_1$ to be well defined 
on the Hilbert space. Consider for example 
$ i \frac{\partial}{\partial y_1^{\prime} } $ acting 
on $ e^{ i y_1^{\prime} \yttp } $. It gives $ - \yttp 
 e^{ i y_1^{\prime} \yttp } $ which does not belong to the Hilbert 
space we defined. The combination $ i 
( \frac{\partial}{\partial y_1^{\prime} } 
-  \frac{ i \yttp}{\yop} \frac{\partial}{\partial \yttp} ) $ 
does map elements in the Hilbert space back to themselves. 
For similar reasons, the appearance of 
$ i(  \frac{\partial}{\partial y_1^{\prime} } 
-  \frac{ i \ytp }{\yop} \frac{\partial}{\partial \ytp} )  $ 
guarantees that the constraint $ \Omega_2 X_1 \Omega_2^{-1} = X_1 $
is satisfied.  The appearance of $ A_2 $ in $X_1$ may seem surprising 
but is necessary to make sure that 
the conjugation of $X_1$ by $ \Omega_3$ does correctly 
reproduce the shift $ - m X_2$ in (\ref{dercom}).  It is also worth noting 
that the change of variables to $y^{\prime} $ coordinates 
is a useful guide in constructing the solution but 
the periodicity conditions are not simple in these coordinates. 
A consequence is that what we might call the gauge 
fields in the primed coordinates, deduced from those 
in the unprimed coordinates are not good operators 
that act in the Hilbert space. For example 
$ A^{\prime}_2 = \frac{1}{y_1^{\prime} } A_2 $ acting 
on the Hilbert space gives functions of the form 
$\frac{1}{y_1^{\prime} } \psi $ which do not belong to the Hilbert 
space. Finally while the above is a relatively simple 
solution including gauge fields, it is not the most general. Just as
in the commutative case, we can also consider $k$-vectors $\psi_k$
acted on by the above operators.
By analogy to the commutative case \cite{ grt} 
 or the case of constant $\theta $  \cite{cds,ho}, 
where the appropriate Hilbert spaces could be generalized to include 
magnetic fluxes we expect similar generalizations here. 

Let us see what happens now in the presence of the harmonic function
$H$ (\ref{noncomm}), and let us restore also the $R_i$ factors. 
The isometries of the metric continue to be the same, $H$ does not
affect the identifications, so we can 
write down the same constraints as before, where now $X^i$ 
correspond to the dimensionless coordinates $x^i$. 
\bea
&&\Omega_1 X_1 \Omega_1^{-1} =X_1 +1 \nonumber\\
&&\Omega_2 X_2 \Omega_2^{-1} =X_2 + 1 \nonumber\\
&& \Omega_3 X_3 \Omega _3 ^{-1} = X_3 +1 \nonumber\\
&& \Omega _3 X_1 \Omega _3^{-1} = X_1 -m X_2 
\label{constra}
\eea
We have only written  the nontrivial relations above.
These have solutions described above. If we rescale $
\tilde{X}_i=r_iX_i$ and correspondingly $\tilde {y}_i =l_s^2y_i/r_i$ for 
the dual variables, then $\tilde{X}_i =il_s^2\tilde{D}_i$,$ \Omega_i =
e^{i\frac{r_i \tilde{y}_i}{l_s^2}}$, 
and one obtains 
\be
[\tilde{y}_2, \tilde{y}_3]
= im \frac{\bar l_s^2 \bar r_1}{\bar r_2 \bar r_3},\;\; \tilde{y}_1= i\tilde{\alpha}\tilde{y}_1,
\;\; [\tilde{y}_1, \tilde{y}_3]=[\tilde{y}_1, \tilde{y}_2]=0~.
\ee
We have thus  obtained the same noncommutativity as from the 
T duality approach (\ref{ncparam}). 
Unfortunately, now there is no independent way to verify the open
string metric and dilaton, which are nontrivial in the presence of 
the harmonic function $H$. 

Still, this fact gives us some useful information about other 
backgrounds. We notice that the 
constraints were not modified by the presence of the D8-brane
background (i.e. by the nontrivial $H$). We can guess that for a general 
massive IIA background, the M theory lift will be again a ``dressing''
of the same space $B(A,R)$ with the same identifications for the dimensionless
$x_i$'s, therefore the constraints (\ref{constra}) are unmodified.
So by the above procedure we will get a \SYM\ on a space with the
same noncommutativity. Again  we will have a nontrivial
metric and dilaton as well as possible other RR fluxes, 
which will have to be derived from the T duality approach. 
The identifications in M theory and 
correspondingly the noncommutativity of the 
D3 brane space are of a topological nature, and so insensitive to local
modifications.

\section{NCSYM action and stability}

In this section we describe the D3 brane action we are getting for the Matrix
model. First let us check that we can put D0
branes at $x=0$ in the background (\ref{twoa}) (and they are
stable). We will also check whether they can stay at nonzero $x$ (in the 
D3 Matrix model, whether we can have a nonzero vev for $X$). 

A probe D0 brane in the background (\ref{twoa}) will have the action
\be
S_{1D0}=  \int dt e^{-\Phi}
 \sqrt{1+ g_{ij}(\dot {X}^i \dot {X}^j)}
\ee
The equations of motion of this action in the background (\ref{twoa})
are
\bea
&&2r_3^2 \frac{d}{dt} [H \dot{X}_3]=2r_1^2 m\dot{X}_2 H^{-1} (\dot{X}_1
+ m X_3 \dot {X} _2)\nonumber\\
&&2\frac{d}{dt} [H \dot{X}] = H' [(\dot{X}^2 +r_3^2 \dot{X}^2_3 +r_2^2 
\dot{X}_2^2 ) -r_1^2 H^{-1}(\dot{X}_1 + mx_3 \dot{X}_2)^2 ]\nonumber\\
&& \frac{d}{dt} [H^{-1} (\dot{X}_1 +mX_3 \dot {X}_2)] =0 \nonumber\\
&& \frac{d}{dt} [ \frac{2mX_3}{H}(\dot{X}_1 + mX_3 \dot{X}_2) + H
r_2^2 \dot{X}_2] =0~.
\eea
It follows that if $\dot{X}_1=\dot{X}_2=\dot{X}_3=0$ and $X$ small (so
that $H\simeq 1$, then the only remaining equation is 
\be
2 \ddot{X} = -H'(\dot{X})^2
\ee
hence the static potential vanishes.

Let us compare this with what happens for the D0-D8 system. There we
have a 1-loop Chern-Simons term $k\int dt (X+A_0), k\in Z$, which 
gives a potential for the D0's. One can calculate it in string theory 
\cite{bgl} or directly from arguments about the supersymmetric quantum
mechanics \cite{lowe1,bss,Kim:1997uv}. But one can understand it from charge
conservation. When a D0 passes through a D8 charge conservation
requires the creation of an elementary string (Hanany-Witten effect),
which will generate a linear potential. In its absence, the D0-D8 is not
supersymmetric and has a linear repulsive potential $V(R)=-T_0 R$. 
Such a Chern-Simons term (and consequently the linear potential) are
absent in the geometric background we consider (\ref{twoa}). 
One can also see this from the D0 worldline
perspective. The CS term of the D0-D8 system 
appeared by integrating out the massive
$(0,8)$ fermion, which is absent from our case. 

Now that we have established that we can have D0 branes at fixed $x$
(and correspondingly D3 branes in the T dual picture), we would like
to describe the action of these D branes in more detail.
The prescription of Myers \cite{myers} 
for the (bosonic part of the) Dirac-Born-Infeld (DBI) action in a 
general background is 
\bea
S_{DBI}&=& -T_p \int d\sigma ^{p+1} {\rm STr}\left( e^{-\phi}
\sqrt{ -{\rm det} ( P [ E_{ab} +E_{ai}(Q^{-1} -\delta)^{ij} E_{jb}] +\lambda 
F_{ab} ) {\rm det} Q^i_j} \right)\nonumber\\
Q^i_j &=& \delta^i_j +i\lambda [X^i, X^j] E_{kj}\nonumber\\
E_{\mu\nu}&=& (g+B)_{ij}
\eea 
and a corresponding Chern-Simons piece. Here the fields are in closed string
variables and if the fields depend on the transverse scalars, the
prescription  is to write the fields as functions of the
adjoint-valued scalars and then take a completely symmetric trace over 
all adjoint indices. 

So the DBI action in our background (\ref{closed}) will be 
\be
S=-T_3\int dt \left(\prod_{i=1}^3 dx_i\right) {\rm STr} \left(e^{-\phi( \tilde{X}^{mn}, x_1)} 
\sqrt { -{\rm det} (g_{ab} ( \tilde{X}^{mn}, x_1)  +B_{ab}( \tilde{X}^{mn}, x_1) 
+\lambda F_{ab})} \right)~,
\ee
here $\tilde X^{mn}$ is the Matrix scalar corresponding to the
coordinate transverse to the D8-brane, and $x_1$ is as defined before.
If we assume that the Seiberg-Witten map continues to hold in the 
presence of the nontrivial $\tilde{X}^{mn}$ (which is not entirely 
obvious, but should probably be true in a Taylor expansion 
for small values of $\tilde{X}^{mn}$), then we get 
\be
S=-T_3 \int dt \left(\prod_{i=1}^3 dx_i\right) {\rm STr} \left(e^{-\tilde{\phi}(\tilde{X}^{mn})}
\sqrt{-{\rm det} (G_{ab}(\tilde{X}^{mn})+ \lambda \tilde{F}_{ab} )} \right)
\ee
where $\tilde{F}$ is the noncommutative field strength. 
Moreover, we saw that the Sen-Seiberg procedure implied that we 
take the Seiberg-Witten limit for noncommutative geometry on the 
D3 action, so we are left with noncommutative \SYM\ with $X$-dependent 
metric and dilaton. 

The D8 brane had 16 supersymmetries, and correspondingly the M theory 
background had also 16 supersymmetries, which means that the D3 brane
action (noncommutative \SYM) will have 8 linearly realized supersymmetries. The 
fermionic field content is the same as for the flat D3 brane, 
but half the supersymmetries are broken by the nonzero $X^{mn}$ and 
the nontrivial noncommutativity.

In the near horizon region (at $x=0$), the D8 background is flat,
so it has 32 supersymmetries. Correspondingly, the DBI action has 
16 supersymmetries if we put $\phi$ and $G_{ab}$ constant (and keep
only the noncommutativity), as in the $\theta$ constant case. 

Let us now examine the star product, since it is nontrivial
(space-dependent). For constant noncommutativity, the noncommutativity 
of the space can be traded for a modified product, the star product,
\be
f\star g= e^{i\theta^{ij} \partial_i \partial_j'}f(x) g(x')|_{x=x'}
\ee
but when $\theta$ is space-time dependent, we have to be more careful.

The first observation we can make is that our $\theta^{ij}$ satisfies
the associativity condition \cite{konts,cs}
\be
\theta^{il}\partial_l\theta^{jk}+\theta^{jl}\partial_l\theta^{ki}+
\theta^{kl}\partial_l\theta^{ij}=0
\label{assoc}
\ee
and so we can define an associative star product. As an aside, 
we have a nonzero $H_{123}$, yet the product is still associative.
This is possible because $\theta_{ij}$ is not invertible in the 
whole space $(1,2,3)$, but just in $(2,3)$
 (if it would be, then associativity and zero H field would be the
 same, see \cite{cs}). Associative star products in the case of 
space dependent $\theta$ can be defined 
with the prescription  given by Kontsevich \cite{konts}. 

The abstract formula is 
\bea
f \star g &=& \sum_n \hbar^n \sum _{\Gamma \in G_n} w_{\Gamma}  B_{\Gamma,
  \theta}(f,g)\nonumber\\
w_{\Gamma}&=& \frac{1}{n! (2\pi)^2n}\int _{H_n} \Lambda_{i=1}^n
(d\phi_{e^1_k} \wedge d\phi_{e^2_k})
\eea
and where explicitly, derivatives which can act either on $f$ and $g$, or
on $\theta$, are contracted with other $\theta$'s. For example, up to
second order in $\theta$ we have 
\bea
f \star g &=& fg + \hbar \theta^{ij} \partial_i f \partial _j g
+\frac{\hbar^2 }{2} \theta^{ij} \theta^{kl} \partial_i \partial_k f 
\partial _j \partial_l g \nonumber\\
&&+\frac{\hbar^2}{3} (\theta^{ij}\partial_j\theta^{kl} )
(\partial_i \partial_k f \partial _l g-\partial_k f \partial_i
\partial_l g) +{\cal O}(\hbar^3)~.
\eea
But in our case we have not only the associativity condition
(\ref{assoc}) but also the more restrictive condition 
\be
\theta^{ij}\partial_j \theta^{kl}=0
\label{starp}
\ee
which implies that there will be no corrections (since derivatives on 
$\theta$ will always appear in the above combination, as the only
object with contravariant indices is $\theta$). Then the Kontsevich
product will be the same as the usual star product, a fact which 
is obvious  in the expanded 
form. The exponential form will also be the same, and we therefore have
\be
f\star g = e^{i\theta^{ij} (x) \partial_i \partial_j'}f(x) g(x')
|_{x=x'}  = e^{i \theta^{ij}(x') \partial_i \partial_j ' }f(x) g(x')|_{x=x'}~.
\ee

As we saw, we can change coordinates by $x_2=x_2'x_1', x_1=x_1'$, and
then ${\theta'}^{23}=-\tilde{\alpha}$, but then the metric is not 
flat anymore. We can however obtain a third form for the star
product. Since in these new coordinates the product is
\be
f \star g= e^{i{\theta '}^{ij}\partial_{x'_i}\partial_{y'_j}}f(x')
g(y')|_{x'=y'}
\ee
by going back to unprimed coordinates we get 
\be
f \star g= e^{i{\theta '}^{ij}\frac{\partial x^m}{\partial {x'}^i}
\frac{\partial y^n}{\partial {y'}^j}\partial_{x_m}\partial_{y_n}}
f(x) g(y)|_{x=y}~.
\ee

As this point it is interesting to observe the similarity of our 
noncommutative theory to the one described by Hashimoto and Sethi 
in \cite{hs}. Moreover, if we take a Penrose-like  limit (infinite
boost, while taking a relevant mass parameter to zero) we obtain 
``half'' their solution. Indeed, 
take an infinite boost in the $x_1$ direction,
and also take $\tilde{\alpha}$ to zero as
\be
x_1\simeq \frac{e^{\epsilon}}{2}({x_1}'+t')=\frac{e^{\epsilon}}{2}{x'}^+
,\;\; \tilde{\alpha}=\frac{e^{-\epsilon}}{2} \tilde{\alpha}'
\ee
and drop the primes. Then the open string variables (\ref{noncomm})
become the flat metric (and constant string
coupling), with $\theta^{23}=-\alpha x^+$ (notice that $H=1$ in this 
limit, since $\tilde{\alpha}\rightarrow 0$). Their solution has also 
$\theta^{-3}=-\alpha x^2$. In these coordinates (with spacetime dependent
noncommutativity), their closed string variables (metric, B field
and dilaton) are 
\bea
ds^2&=& [-2dx^+dx^- +\frac{R^2(dx_2^2+dx_3^2)}{R^2+(x^+)^2}]-
\frac{x_2^2(dx^+)^2}{(x^+)^2}+\frac{2x_2x^+dx_2dx^+}{R^2+(x^+)^2}\nonumber\\
B&=&[\frac{Rx^+ dx_2\wedge dx_3}{R^2+(x^+)^2}]-\frac{Rx_2 dx^+\wedge
  dx_3}{R^2
+(x^+)^2}\nonumber\\
e^{\phi}&=&g_s\sqrt{\frac{R^2}{R^2+(x^+)^2}}
\eea
where the terms in brackets correspond to  the ``Penrose'' limit of
our solution. In these coordinates, their open string metric is flat and
the open string dilaton constant, just as in our case. 
So we are obtaining ``half'' the solution in \cite{hs}, which seems to 
suggest that both are part of a 1-parameter set of solutions. 

Another observation is that in \cite{hs}  there is also a 
transformation of coordinates which makes $\theta^{ij}$ constant, namely
\bea
x^+&=& {x'}^+\nonumber\\
x_2&=& x_2'{x'}^+\nonumber\\
x^-&=& {x'}^- +\frac{1}{2} {x'}^+x_2^2
\eea
whereas for the ``Penrose'' limit of our solution it is just
\be
x^+={x'}^+ , \;\;
x_2=x'_2{x'}^+ ~.
\ee

However, in their case (\ref{starp}) is not satisfied, while
(\ref{assoc}) is still satisfied, so in their case the Kontsevich 
product is different from the usual star product, even though there 
is a coordinate transformation which makes $\theta$ constant. 

Finally we  note that an example of spacetime dependent
noncommutativity has been analyzed in \cite{dn}, and the
Seiberg-Witten analysis still holds (even though $\theta$ is varying).

\section{Spectrum of states}

Now we take a step toward deriving the duality between the 
noncommutative \SYM\
on (\ref{noncomm}) and massive IIA in the D8 background, by studying
the spectrum of BPS states. Type IIB string theory in ten dimensions
can be obtained by compactifying M-theory on a 2-torus of vanishing
area, but fixed complex structure.
In this case the 
Sethi-Susskind \cite{ss} and Banks-Seiberg \cite{bs} 
constructions gave evidence
for the duality. We will follow the Sethi-Susskind construction, which
is defined in 3+1 dimensions, setting it up so that 
we can go smoothly to our case. The IIB Matrix model is 2+1 \SYM\, which 
has naturally an $SO(7)$ invariance, but the claim is that at strong
coupling it develops an $SO(8)$ invariance (which is consistent with the 
supersymmetry algebra and is the maximal $R$ symmetry). The easiest way to see it 
\cite{ss} is to go to 3+1d \SYM\ and use
electric-magnetic duality. There we have only an $SO(6)$ manifest
invariance (6 scalars), which will be enhanced to $SO(8)$. 
In our case we naturally have 3+1d \SYM, so
it should be our starting point. In the massless case ($m=0$), we still have
an $SO(6)$ (6 scalars) enhanced to $SO(8)$, but in the massive case we
have an $SO(5)$ (the scalar $X$ vev corresponding to the direction
transverse to the D8 brane in IIA - is special), which should be
enhanced to $SO(7)$.

Let us then set up 3+1d \SYM\ for our use. 
The \SYM\ lives on a dual torus of lengths $\tilde{R}_1, \tilde{R}_2,
\tilde{R}_3$.
Sethi and Susskind have $R_2, R_3\rightarrow 0, 
R_1\rightarrow \infty$. The mass of a membrane on the shrinking torus 
$R_2, R_3$ is identified via the M theory-IIB duality 
with the momentum mode on another direction $Y$ in IIB,
\be
\frac{R_2R_3}{l_P^3}=\frac{1}{R_Y}
\label{wrap}
\ee
with the limit $R_Y$ to infinity, and we set $R_Y=R_1$. 
By the above formula, we see that $R_Y=
l_s^2/R_2$ 
(if $R_3$ is the M theory direction), and so as we said 
$R_Y$ is the
extra transverse direction in the lightcone IIB theory which appears
when the M theory torus shrinks to zero size. To obtain $SO(8)$
invariance, we indeed need to choose $R_Y=R_1=R_{\perp}$, so that 
all the IIB lightcone coordinates, $X_1, X_Y, X_4,...X_9$ have length
$R_{\perp}$.   Then the 3+1d \SYM\ coupling,
\be
\tilde{g}_{YM}^2 = \frac{l_P^3}{R_1R_2R_3}=1
\label{coup}
\ee
so we are at the self-dual point, and we have electric-magnetic
duality. As usual by the Sen-Seiberg procedure, $\bar{M}$ theory 
was introduced in order to show massive string degrees of freedom
decouple from the \SYM, but the T dual \SYM\
variables depend only on M theory quantities. 

The electric flux along $\tilde R_1$ corresponds to the
momentum conjugate to $X_1$ under T duality $ \partial A_1 \rightarrow
\partial X_1$, and so it goes together with the other momenta to 
increase $SO(6)$ to $SO(7)$ invariance. Because of electric-magnetic
duality 
however, $SO(7)$ becomes $SO(8)$.

In our case, when the 3d space $B$ in M theory  shrinks to zero size, 
we have 2 extra transverse lightcone coordinates appearing in type IIB,  
\be
\frac{R_1R_2}{l_P^3}=\frac{1}{R_{Y_2}} {\rm and} \;\; 
\frac{R_1R_3}{l_P^3}=\frac{1}{R_{Y_3}}
\label{wrapp}
\ee
and so $SO(8)$ invariance (in the massless case) 
should be recovered when $X_{Y_2}, X_{Y_3}, X_4,..., X_9$ have the
same length, 
\be
R_{Y_2}=R_{Y_3}=R_{\perp}
\ee

For Sethi and Susskind, 
the magnetic flux $F_{23}$ in the D2 theory (and correspondingly in 
the D3 theory as well) was T dual to wrapping number of membranes,
which by the M-IIB duality (and 9-11 flip) was identified with 
momentum on $R_Y$ (see (\ref{wrap})). 
For our case, the invariance we seek is with the momentum in 
$R_{Y_2}$ and $R_{Y_3}$, which corresponds in the D0 theory 
 by (\ref{wrapp}) to wrapping number on 12 and 13 respectively. By T
 duality, in the D3 theory, this is magnetic flux $F_{12} $ and
 $F_{13}$. 

So let us see the $SO(8)$ invariance in the 2 cases from the YM energy.   
The energy of magnetic fluxes and electric fluxes is deduced from 
\bea
&&\int dx^2 dx^3 {\rm tr} F_{23} = n_m^{23}\nonumber\\
&& \int dx^2 dx^3 {\rm tr} F_{01} =n_e^{23} g_{YM}^2 
\eea
(for magnetic flux on 23 and electric flux on 1) so that 
\bea
&& {\rm tr}F_{23}=\frac{n_m^{23}}{\tilde{R}_2\tilde{R}_3}\nonumber\\
&& {\rm tr}F_{01}=\frac{n_e^{23} g_{YM}^2}{\tilde{R}_2\tilde{R}_3}~.
\eea
If we add momentum modes on $x_1, x_2, x_3$ ($p^i=n^i/\tilde{R}_i$),
 the energy 
\be
E=\frac{1}{2g_{YM}^2}\int dx^1 dx^2 dx^3 [{\rm tr}F_{0i}^2 +\sum_{j<k}
{\rm tr}F_{jk}^2 +
(\partial_{\mu} X^i)^2 ]
\ee
becomes (using that ${\rm tr}_{U(N)}F_{\mu\nu}^2={\rm tr}_{SU(N)}F_{\mu\nu}^2
+1/N ({\rm tr} F_{\mu\nu})^2$ and concentrating on the U(1) piece)
\bea
N E &=& \frac{\tilde{R}_3}{2\tilde{R}_1\tilde{R}_2}[ \frac{n_{m12}^2 }
{\tilde{g}_{YM}^2} 
+ n_{e12}^2 \tilde{g}_{YM}^2 ] +\frac{\tilde{R}_1}{2\tilde{R}_2\tilde{R}_3}
[ \frac{n_{m23}^2 }{\tilde{g}_{YM}^2} 
+ n_{e23}^2 \tilde{g}_{YM}^2 ]+ \nonumber\\
&&\frac{\tilde{R}_2}{2\tilde{R}_1\tilde{R}_3}[ 
\frac{n_{m13}^2 }{\tilde{g}_{YM}^2} 
+ n_{e13}^2 \tilde{g}_{YM}^2 ]+ \sqrt { 
( \frac{n_1}{\tilde{R}_1})^2 + ( \frac{n_2}{\tilde{R}_2} )^2 
+ ( \frac{n_3}{\tilde{R}_3} )^2 } 
\label{energy}
\eea
This formula is also in accord with \cite{ho,pioline}.
The elementary excitations of the theory are the momentum modes $n_i$,
but  the Matrix theory prescription tells us to look at the energy 
of excitations on the moduli space, in other words for excitations with
energy much smaller than that of momentum modes. 

In the Sethi-Susskind case ($\tilde{g}_{YM}=1$), 
the smallest elementary excitation (momentum mode) is 
of order $1/\tilde{R}_{2,3}$ ($\tilde{R}_1 \ll \tilde{R_{2,3}}$), and
the 12 and 13 fluxes have energy much bigger than that, whereas  the
23 fluxes have much smaller energy, so they should be thought of as moduli.

In our case, ($\tilde{g}_{YM}\to \infty$, $\tilde{R}_1 \gg \tilde{R}_{2,3}$),
the smallest elementary excitation is of order $1/\tilde{R}_1$, and
the 23 fluxes have energy much bigger than that. For the 12 and 13
fluxes, choosing $\tilde{R}_2=\tilde{R}_3$ (as we have seen we need in 
(\ref{wrapp})), the prefactor (energy scale) 
of the fluxes is also $1/\tilde{R}_1$,
as for the momentum modes. However, since $\tilde{g}_{YM}$ is infinite, the
electric fluxes have energy much larger than the momentum modes,
whereas the magnetic fluxes have energy much smaller than the momentum 
modes, so are real moduli. 

Now let us see what should we compare the energy of those moduli
against. The energy on the moduli space of the 6 scalar fields is 
\be
N E=\frac{{\vec{p}_{\perp}}^2}{2M_0}=\frac{{\vec{n}}^2}
{2\bar{R}_{\perp}^2M_0}=\frac{{\vec{n}}^2}{2M_0 R_{\perp}^2}\frac{l_P^2}{
\bar{l}_P^2}
\label{nonrel}
\ee
where we have put the transverse space in a box of size $R_{\perp}$,
to be equated with $R_{Y_2}=R_{Y_3}$ as above, and $M_0$ is the mass of 
those moduli, but we have taken into account that we are calculating
energies in the $\bar{M}$ theory.
$M_0$ comes from the fact that we are really expanding the 
DBI action of the D3 in order to get (\ref{nonrel}).  But then
\be
M_0=V_pT_p =\frac{\tilde{R}_1\tilde{R}_2\tilde{R}_3}{\tilde{g}_s
  \bar{l_s}^4 }=\frac{1}{\bar{g}_s \bar{l}_s} \rightarrow 
NE=\frac{{\vec{p}_{\perp}}^2}{2M_0}= \frac{ {\vec{n}}^2\bar{g}_s
  \bar{l}_s}{2R_{\perp}^2
  }\frac{l_P^2}{\bar{l}_P^2}=R\frac{{\vec{n}}^2}{2R_{\perp}^2}
\ee
Here $\bar{l}_s$ is the string length for the $\bar{M}$ theory. 
Then in the Sethi-Susskind case we have the energy of the moduli
(using (\ref{coup}))
\be
E= \frac{l_P^2}{2N}\frac{\bar{l}_s \bar{g}_s}{\bar{l}_P^2}
 [\frac{n_{m12}^2 +n_{e12}^2}{{R}_1^2}
+\frac{{\vec{n}}^2}{R_{\perp}^2}]
=\frac{R}{2N}[\frac{n_{m12}^2 +n_{e12}^2}{{R}_1^2}
+\frac{{\vec{n}}^2}{R_{\perp}^2}]
\label{sesu}
\ee
which is $SO(8)$ invariant if we put $R_y=R_1=R_{\perp}$, and in our
case (with $m=0$) we get (using that $\tilde{g}_{YM}^2=l_s^2/(R_2R_3), 
\tilde{R}_1=l_s^2/R$ and (\ref{wrapp}))
\be
E= \frac{R}{2N}[\frac{n_{m,13}^2}{R_{Y_2}^2}+\frac{n_{m,12}^2}{
R_{Y_3}^2}+\frac{{\vec{n}}^2}{R_{\perp}^2}]
\label{us}
\ee
which is $SO(8)$ invariant if we put $R_{Y_2}=R_{Y_3}=R_{\perp}$. 
We notice that the formulas (\ref{sesu}) and (\ref{us})  are 
exactly what we expect from supergravity and from the BFSS model \cite{bfss} 
for the free supergravitons. Of course it would be more interesting 
to derive the interaction piece.

Finally, what happens in the massive case $m\neq 0$ (in the $X=0$ sector, which
has still 16 supersymmetries)?  As we mentioned, one of
the scalars ($X$) corresponds to the direction transverse to the D8 brane,
so we have manifest $SO(5)$ invariance of the scalars which should be
lifted to a $SO(7)$ invariance, of the lightcone string theory in the 
D8 background. So the above formulas should apply to 5 of the scalars,
but not to the $X$ direction.

The noncommutative  \SYM\ is obtained by replacing the usual product
with the star product. 
To first order in $\theta$, the action (see \cite{sw}, equation
4.27) is 
\bea
S&=&\int [ F_{ij}F_{mn}G^{im}G^{jn}(1-1/2 F_{ij}\theta^{ij}) 
-2\theta^{kl}F_{ki} F_{lj} F_{mn}G^{im}G^{jn}]\nonumber\\
&=& \int[F_{ij}^2(1+3F_{23}\theta^{23})]
\eea
and we see that if $F_{23}=0$, the action is unmodified, and so the 
energy formula is unmodified as well, as expected. The higher order
terms will just contain terms with derivatives of $F$, and so a constant 
$F_{23}$ and $F_{13}$ will still be a solution, and the energy formula
of the magnetic fluxes will again be unmodified. As for the momentum 
modes $\vec{p}_{\perp}$, 
they are modes on the moduli space, not in worldvolume, so
there is no reason for their energy to be modified. 
We can easily verify that there are no $\theta$ corrections
to the energy in our case by applying the general formula in
\cite{ho,pioline} to the moduli space of our theory.

As for the action of the elementary strings, that is easy to
understand. In the Type IIB case, \cite{bs} considered the limit of
small coupling, when $R_3 \ll R_2 \rightarrow \tilde{R}_2 \ll
\tilde{R}_3$, so Yang-Mills moduli space excitations occur only 
in the 3 direction. Then the string action is just the sigma model action on 
the moduli space, with worldvolume given by time and the 3 direction.
In our case, there is no need to take small coupling, since already
$R_1 \ll R_{2,3}\rightarrow \tilde{R}_1 \gg \tilde{R}_{2,3}$, and so 
Yang-Mills moduli space excitations already occur just in the
worldvolume 1 direction, and the string action is again the sigma model. 

What about the supergravity mass? Massive IIA supergravity
\cite{romans} has a massless graviton, a dilaton and an antisymmetric
tensor $B_{\mu\nu}$ which acquire a mass proportional to $m$, a 
massless 3-form $A_{(3)}$, and massive fermions (gravitino and spin 
1/2). The 1-form $A_{(1)}$ is gauged away, since it appears in the 
combination $F_{(2)} +mB_{(2)}$. In the D8 brane background, one has
to study the wave equation for each field. The massless fields 
(graviton and $A_{(3)}$) can still have a constant wavefunction in the 
$x$ direction (transverse to the D8), and then 
\be
p^2=-2p^+p^-+\vec{p}_{\perp}^2=0
\rightarrow E=p^-=\frac{\vec{p}_{\perp}^2}{2p^+}
\ee
 (where $p$ is the momentum along the D8)
which reproduces  (\ref{us}). A nontrivial wavefunction in the $x$
direction implies that $p^2 \neq 0$, and correspondingly an extra term
in the energy. 

For one of the massive fields, we have to study the wave equation 
in the D8 background. The Einstein metric is 
\be
ds_E^2=H^{1/8} dy_i^2 + H^{9/8} dx^2
\ee
which means that the wave equation for a scalar of mass
$aM $ in this background (the dilaton is such a scalar), with $a$
a constant, is 
\be
(\Box -a^2M^2)\phi =0 \Rightarrow 
(\partial_i^2 +H^{-1}\partial_x^2 -a^2M^2)\phi =0
\ee
and then for a separated solution,
\be
\phi =e^{ip_iy_i}\phi(x)
\ee
we get 
\be
(\partial_x^2 -p^2(1+M|x|) -a^2M^2(1+M|x|)^{9/8})\phi (x)=0~.
\ee
Notice that this equation does not admit a constant wavefunction, 
since $\phi(x)=c, -p^2-a^2M^2=0$ is not a solution.
At large x, it has the asymptotic solutions
\be
\phi(x)=e^{\pm \frac{16}{25}a (Mx)^{25/16}}
\ee
so we can keep only the decaying solution
and at small x it becomes a combination of the oscillatory solutions
\be
\phi (x)= e^{ix\sqrt{-p^2-a^2M^2}}
\ee
(for $p^2< -a^2M^2$).
So from matching the wavefunction and its derivative over $x=0$ 
we get a condition on $p^2$, which will be that the coefficient of 
the $sin$ should be zero, which will imply a quantization condition,
of the type
\be
p^2=-M_n^2(a,M)~.
\ee
The same type of analysis should hold for every massive 
field in supergravity, so in general we will get a formula for the 
lightcone energy of the type 
\be
E=\frac{\vec{p}_{\perp}^2+ M_n^2(a,M)}{2p^+}~.
\ee
Correspondingly, we expect to find in \SYM\ that various moduli
have such an additional mass term, which will depend on the detailed
structure of the interactions permitted by the 8 linearly realized
supersymmetries. 
However, these massive 
moduli will appear only when we look at nontrivial wavefunctions 
for the \SYM\ scalar $X$ (other than $X$=constant and small), so it is
hard to analyze. The moduli with trivial wavefunctions in the 
$X$ direction will correspond to the massless supergravity modes, 
and as we saw, they have the right lightcone energy. Moreover, even 
if we would find the massive \SYM\ moduli, on the supergravity side it is 
also hard to get any results  (although one could of course use
numerical methods to find the mass terms).  

But we can make one observation. On the supergravity side, all the 
fermions are massive, so we expect that also fermionic \SYM\ moduli 
will have a mass term in the energy. One hint that this might happen 
is that we expect the D3 brane fermions to have a worldvolume mass 
term of the type $aM$. Indeed, although we do not know how to write 
down the D3 fermionic action for a general supergravity background, 
we know that in some backgrounds (like super-coset manifolds), the 
kinetic term for the fermions is of the type $\bar{\psi} {\cal
  D}\psi$, where ${\cal D}$ is the spacetime Killing spinor operator 
pulled back on the worldvolume \cite{kallosh}. The kinetic term then
 contains a term of the type
\be
\alpha 'H_{\mu\nu\rho}\bar{\Psi}\Gamma^{\mu\nu\rho}\Psi
\ee
which would imply a mass proportional to  $\bar{l}_s^2 H_{abc}$ (flat
indices). But for our closed string background in the 
Seiberg-Witten limit (\ref{closed}), $B_{23}=1/(\tilde{\alpha} x_1)$
and $g_{22}=g_{33}=(\bar{l}_s^2/(\tilde{\alpha} x_1))^2$, and so 
the fermion mass will be proportional to 
\be
\bar{M}=\frac{\tilde{\alpha}}{\bar{l}_s^2}= \frac{m
  \bar{r}_1}{\bar{r}_2\bar{r}_3}=\frac{mr_1}{r_2r_3}\frac{l_P}{\bar{l}_P} 
=M \frac{l_P}{\bar{l}_P}
\ee
with M being the supergravity mass. Then the moduli mass will be 
$\bar{M}_n(a, \bar{M})=M_n(a,M) l_P/\bar{l}_P$ and if we would get an 
energy $\bar{M}^2_n(a, \bar{M})/2NM_0= M_n^2(a,M)/2p^+$, it would be
as desired. 
It would be of course very interesting to see whether one can recover
all the supergravity mass terms for the lightcone energy, but as we 
saw, the analysis looks quite difficult.

\section{Holographic dual}

Let us try to write down the holographic dual of our noncommutative 
\SYM\ defined on (\ref{noncomm}) in the spirit of the AdS/CFT correspondence. 
We have to write down a solution for D3 branes in the 
closed string background 
(\ref{closed}) corresponding to (\ref{noncomm}) 
and then take a decoupling limit. It turns out however
to be easier to start with D1 branes in the background (\ref{nsfive})
and then make two T dualities. Indeed, 
as we saw, the background (\ref{nsfive}) corresponds to NS5
branes smeared over the directions 1,2,3. We have
to put D1 branes at $x=0$ along time and $x_1$. But this background 
is one of D1 ending on NS5's, smeared over the D1 direction, as well
as 2 others, transverse to both NS5 and D1. The S dual of this (type
IIB) configuration is F1 ending on D5, which we know that exists. Then
the original IIA metric is D0's parallel to KK monopoles, and after
$T_1$ and $ T_2$ we have D2 ending  on KK monopoles, and finally after $T_3$ 
we have D3 ending on an unusually T dualized KK monopole (an 8
dimensional worldvolume). 

The solution for D1 ending on NS5's, depending only on the coordinate 
$x$ can be found pretty easily, namely
\bea
ds^2 &=& -dt^2 H_1^{-1/2} +H_1^{1/2} d\vec{\sigma}^2_5 + H H_1^{-1/2} 
dx_1^2/r_1^2 + HH_1^{1/2} (dx^2 +r_2^2 dx_2^2 +r_3^2
dx_3^2)\nonumber\\
B_{12} &=& m x_3\nonumber\\
e^{\phi}& =& \frac{e^{\phi_0}}{ r_1} H^{1/2} H_1^{1/2}~.
\label{done}
\eea
T dualizing on $T_2$ we get 
\bea
ds^2 &=& -dt^2 H_1^{-1/2} +H_1^{1/2} d\vec{\sigma}^2_5 + H H_1^{-1/2} 
dx_1^2/r_1^2 + HH_1^{1/2} (dx^2  +r_3^2dx_3^2) \nonumber\\
&&+ \frac{H^{-1}
  H_1^{-1/2}}{r_2^2}(dx_2+ mx_3 dx_1)^2 \nonumber\\
e^{\phi} &=& \frac{e^{\phi_0}}{ r_1r_2} H_1^{1/4} ~.
\label{donee}
\eea
And finally, after the coordinate transformation (\ref{coordtr1})
and $T_3$ T duality, we get (putting back the $l_s$ dependence )
\bea
ds^2 &=&l_s^4[ -dt^2l_s^{-4} H_1^{-1/2} +H_1^{1/2} l_s^{-4}
d\vec{\sigma}^2_5 + H H_1^{-1/2} 
dx_1^2/r_1^2 + HH_1^{1/2} l_s^{-4}
dx^2  \nonumber\\
&&+ \frac{H^{-1}
  H_1^{-1/2} }{1+ l_s^4(\frac{r_1}{r_2r_3})^2 \frac{m^2 x_1^2/r_1^2 }{H^2
    H_1}} (dx_2^2/r_2^2 +dx_3^2/r_3^2)] \nonumber\\
B_{23}&=& -l_s^4\frac{m x_1 }{r_2^2 r_3^2 H^2 H_1} \frac{1}{1+ l_s^4
(\frac{r_1}{r_2r_3})^2 \frac{m^2 x_1^2/r_1^2 }{H^2    H_1}}
\nonumber\\
e^{\phi} & =& \frac{l_s^3e^{\phi_0}}{ r_1r_2r_3 H^{1/2} } (1+ l_s^4                                                            
(\frac{r_1}{r_2r_3})^2 \frac{m^2 x_1^2/r_1^2 }{H^2    H_1} )^{-1/2} ~.
\label{holodual}
\eea
However we need to generalize this to the fully localized solution,
where the D3-branes are not smeared over the transverse directions.

The first thing we can do is to look in the near core region. In the
near core region, $H$ is constant ($\simeq c$), and then there is no 
obstruction for making the harmonic function $H_1$ depend on all its 
transverse coordinates. Indeed, since $H$ is constant, we can ask to
find a D1 brane solution in the corresponding flat background, and 
that is just the usual D1 brane with a nontrivial $B$ field, i.e.
(\ref{done}), with $H=c$ and $H_1(\vec{\sigma}_5, x_2, x_3, x)$ 
the usual harmonic function. Then after the two T dualities one gets
the solution (\ref{holodual}), where $H=c$ and 
\be
H_1\simeq 1+ \frac{4\pi g_s N {\alpha '} ^2}{ (\vec{\sigma}^2 +cx^2)^2
}\simeq 1+ \frac{ 4\pi g_s N {\alpha '} ^2}{ \vec{\sigma}^4}
\label{harmonic}
\ee
where in the last line we have used the fact that we work near $x\simeq
0$. 

Let us now  derive the equation for the 
full solution (outside the core).  Partially localized intersections,
where brane 1 with harmonic function $H_1$ 
lives on $t, \vec{w}, \vec{x} $, and brane 2 with harmonic function
$H_2$ lives on 
$t, \vec{w}, \vec{y}$, with overall transverse space $\vec{z}$,
are written in terms of harmonic functions $H_1$ and $H_2$ in the
usual way, except that now $H_1$ and $H_2$ 
satisfy the equations (e.g. \cite{youm}, \cite{lupo})
\bea
\partial_z^2 H_1(z, y) + H_2(z) \partial_y^2 H_1(z, y)=0, \partial_z^2
H_2=0 \;\; {\rm or}\nonumber\\
\partial_z^2 H_2(z, x) + H_1(z) \partial_x^2 H_2(z, x)=0, \partial_z^2 
H_1=0~.
\eea
In other words, we delocalize one brane (say brane 2) over the
worldvolume coordinates of the other brane (1), 
and then $H_1$ is harmonic  (obeys the Laplace equation) 
in the background of brane 2. This is true for any kind of branes, but
in particular \cite{lupo} has derived explicitly this equations for the 
11d intersection of M2 and M5 (over a string). This intersection is
related to our D1-NS5 solution as follows. Dimensionally reduce to
type IIA on the common string, to an F1-D4(0) solution, T dualize to
IIB on a transverse direction to a F1-D5(0), and then S dualize to 
D1-NS5(0). 

In our case, the harmonic function $H$ is delocalized over the D1, that
is over $x_1$, as well as over $x_2, x_3$, over which we need to T
dualize, and $H_1$ is delocalized over $x_2, x_3$.
 So the full solution is given by (\ref{done}), where $H_1$
satisfies the equation
\be
[\partial_x^2 +H(x)\partial_{\vec{\sigma}}^2 ]H_1 (\vec{\sigma},
x)= Q \delta (\vec{\sigma}) \delta (x)
\label{harmeq}
\ee
where we have put explicitly the source term $Q=16\pi^4 g_s N
(\alpha')^2$. 
Then also (\ref{donee}) and (\ref{holodual}) are the corresponding
T dual solutions. 
We notice that near the core $x=0$, $H\simeq c$, so the solution is 
indeed (\ref{harmonic}).

In order to solve (\ref{harmeq}), we separate variables, by writing
\be
H_1(\vec{\sigma}, x)= 1+ \int \frac{d^5p}{(2\pi)^5}e^{i\vec{p}\cdot
  \vec{\sigma}} H_{1,p}(x)=1+ \frac{1}{8 \pi^3}\frac{1}{r^2} \int
_{0}^{\infty} dp p^2 (\frac{\sin(pr)}{pr}-\cos(pr)) H_{1,p}(x)
\ee
and get the equation
\be
H''_{1,p}(x)-p^2 H(x)H_{1,p}(x)= Q \delta (x)~.
\ee
By putting $H=c + m|x|$  (we will keep this form for now and replace 
it later with $c=1$ and $m=\tilde{\alpha}/l_s^2$) and 
\be
\bar{x}= (\frac{p}{m})^{2/3}(c +m |x|)
\ee
we get the Airy equation 
\be
\frac{d^2 H_p}{d\bar{x}^2}-\bar{x} H_p(\bar{x})= Q m^{-1/3} p^{-2/3}
\delta (\bar{x}- c (p/m)^{2/3})
\ee
which has solutions in terms of the Bessel functions $I_{1/3}$ and 
$K_{1/3}$. We choose $K_{1/3}$ which decays exponentially at infinity, 
and get
\be
H_p(\bar{x})= c_p \bar{x}^{1/2} K_{1/3} \left(\frac{2}{3} \bar{x}^{3/2}\right)~.
\ee
The coefficient $c_p$ can be fixed by matching with the normalization 
of the $\delta$ function source. We get
\be
c_p = \frac{Q \sqrt{c}}{2p^{1/3} m^{2/3} [K_{1/3} ( \frac{2}{3}
  \frac{p}{m} c^{3/2})- \frac{p}{m}c^{3/2}K_{4/3} (\frac{2}{3}
\frac{p}{m} c^{3/2})]}~.
\ee
Therefore the final formula for the harmonic function is 
\be
H_1(r,x)=1+\frac{Q\sqrt{c}}{8\pi^3 m^{2/3}}\frac{\beta^{1/3}}{r^2}
\int dp p^2 \frac{(\frac{\sin(pr)}{pr}-\cos(pr)) K_{1/3} (\frac{2}{3} \beta p)}
{K_{1/3} ( \frac{2}{3}
  \frac{p}{m} c^{3/2})- \frac{p}{m}c^{3/2}K_{4/3} (\frac{2}{3}
\frac{p}{m} c^{3/2})}
\label{harmonicf}
\ee
with
\be
\beta =  \frac{(c+m|x|)^{2/3}}{m}~.
\ee
So we have found the full solution for the D3 branes in the
background. 

We can now write down the decoupling limit for the holographic dual
in the near core ($x=0$), namely 
\be
\alpha ' \rightarrow 0, U =\frac{|\vec{\sigma}|}{\alpha '} ={\rm
  fixed}, \; X=\frac{x}{\alpha '}={\rm fixed}, \; g_s N = \lambda= {\rm fixed} 
\ee
but we have to supplement it with 
\be
r_i \rightarrow 0, \tilde{y}_i = \frac{l_s^2 x_i}{r_i}= {\rm fixed}, 
\;\; \tilde {\alpha} = \frac{m  r_1l_s^2}{r_2 r_3}={\rm fixed}
\ee
and then we have the holographic dual
\be
ds^2 = \alpha ' [ \frac{U^2}{\sqrt{\lambda}}(-dt^2 +d\tilde{y}_1^2 
+\frac{d\tilde{y}_2^2+d\tilde{y}_3^2}{1+
\frac{\tilde{\alpha}^2 \tilde{y}_1^2
    U^4}{\lambda}})
+\frac{\sqrt{\lambda}}{U^2}(dX^2 +dU^2 +U^2 d\Omega_4^2)]~.
\label{holodualc}
\ee
This metric is then dual to \SYM\ with
\be
[\tilde{y}_2, \tilde{y}_3]=i\tilde{\alpha}\tilde{y}_1, 
\;\; ds^2 = -dt^2+ d\tilde{y}_2^2 +d\tilde{y}_1^2 +d\tilde{y}_3^2~.
\ee
We note that the holographic dual in the near core region 
(\ref{holodualc}) is just what we would have
expected from the usual noncommutative case \cite{hi,mr},
 with holographic dual
\be
ds^2 =\alpha ' [ \frac{U^2}{\sqrt{\lambda}}(-dt^2 +dy_1^2 +\frac{
dy_2^2 +dy_3^2}{1+\frac{\Delta^4 U^4}{\lambda}})+\frac{
\sqrt{\lambda}}{U^2}(dU^2 +U^2 d\Omega_5^2)]
\ee
and $\Delta^2= \theta^{23}$. 

To get the full holographic dual, since (remembering 
just for the purpose of next formula that what we call 
$l_s$ is really $\bar{l}_s$, whereas $l_s^A$ still appears in $H$
and also $m$ denoting the integer= D8 number)
\be
H= 1+ \frac{m g_s^A}{l_s} |x| =1 +\frac{m g_s^A \bar{l}_s^2}{l_s} |X|
=1+\tilde{\alpha}|X|
\ee
we replace $c=1$, $m=\tilde{\alpha} /l_s^2$, $Q=16\pi^4 g_sN l_s^4$, together 
with the rest of the limit into (\ref{harmonicf}), and rescaling
the integration variable as $p=P/l_s^2$ we get 
\be
H_1 (r,x)\simeq \frac{h_1(U,X)}{l_s^4}
\ee
where
\be
h_1(U,X)= \frac{2\pi g_s N}{\tilde{\alpha}^{2/3}}
\frac{\bar{\beta}^{1/3}}{U^2}\int_0^{\infty} dP P^2 \frac{(\frac{
\sin(PU)}{PU}-\cos(PU))
K_{1/3}(\frac{2}{3} \bar{\beta} P)}{K_{1/3}
(\frac{2}{3}\frac{P}{\tilde{\alpha}}) -\frac{P}{\tilde{\alpha}}
K_{4/3}(\frac{2}{3}\frac{P}{\tilde{\alpha}})}
\ee
and
\be
\bar{\beta} = \frac{(1+ \tilde{\alpha}|X|)^{2/3}}{\tilde{\alpha}}~.
\ee

Then the full holographic dual is 
\be
ds^2=\alpha ' [ h_1^{-1/2}(U, X)(-dt^2 +Hd\tilde{y}_1^2+
H^{-1}\frac{d\tilde{y}_2^2+d\tilde{y}_3^2}{1+
\frac{\tilde{\alpha}^2 \tilde{y}_1^2
    U^4}{\lambda}}) +h_1^{1/2}(HdX^2+dU^2 +U^2 d\Omega_4^2)]~.
\ee
We note that the Seiberg-Witten limit is a subset of the holographic 
limit, as it should be.

\section{Conclusions and Discussion}

We have proposed a new nonperturbative formulation of massive Type IIA
string theory in terms of a noncommutative Yang-Mills theory with
space dependent noncommutativity parameter. There remains much to
study. In particular, it would be very interesting to construct in
more detail the interaction terms in the action, the energies of 
physical excitations and to study the
S-duality properties of this noncommutative gauge theory.
A more direct derivation of the non-commutative 
Yang Mills in section 5 starting from the solution of the 
Matrix Theory constraints of section 4, using information about the 
action of zero branes in the curved space of the twisted torus 
will be useful. In fact it may be easier to try and guess
the form of the zero brane action which would lead to the 
actions in section 5, using the $X$ matrices constructed in 
section 4. Progress in these directions is likely to also 
be useful in flux compactifications since T-duality of the twisted 
torus gives a background with $H$-flux as discussed in section 3. 
These compactifications offer promising avenues toward the problem 
of fixing moduli in string  phenomenology  \cite{kstt}. 

We note that we have described massive IIA theory in terms of a matrix
model of D3 branes with noncommutativity, a theory which has a
holographic dual. As a limit, massless IIA theory is described by a
matrix model of D3 branes, which is dual to string theory in $AdS_5
\times S_5$. But there are two things we should observe:

1)The D3 branes are on a torus, which translates in making
identifications in $AdS_5$ (in Poincare coordinates, $ds^2= y^2 (-dt^2
+d\vec{x}^2)+dy^2/y^2$, and the $\vec{x}$ coordinates are identified 
on a torus).

2)There are different observables in the D3 brane theory which
describe flat space IIA string theory and $AdS_5 \times S_5$ 
string theory. For $AdS_5\times S_5$, we look at gauge invariant
observables in the D3 brane theory, whereas for the IIA matrix model 
we look at wavefunctions on the moduli space, thereby  spontaneously breaking gauge
invariance. Holographic duals in the context of 8-brane solutions 
 have also been discussed recently in \cite{singh}.

We comment  on the relation of this 
construction to Type IA string
theory \cite{pw}, where D8-branes and O8-planes coexist. The massive
Type IIA physics is recovered by focusing on the local physics
between a pair of separated D8-branes, or equivalently, by sending the
D8-branes and O8-planes off to infinity. There exists a Matrix
proposal for the complete nonperturbative Type IA system
\cite{lowe2,motl2,Rey:1997hj,Kabat:1997za} which is related by S-duality to the $E_8\times
E_8$ heterotic string. It would be interesting to recover the
noncommutative theory described in this paper by integrating out
degrees of freedom in these heterotic Matrix models.

As we mentioned in section 2, a generalized Scherk-Schwarz
reduction based on a scaling symmetry of the equations 
of motion gives a ten dimensional supergravity which 
has de Sitter solution \cite{chamblin}. 
It was observed in  \cite{lalupo} that these
can be viewed in terms a Euclidean radial reduction from M Theory. 
This suggests 
that a Matrix Model could be found by generalizing the dimensional 
reduction methods of Matrix Theory that we have used to 
radial reductions. This is of course a non-trivial generalization 
since the spacetime of M-Theory, and hence a  Euclidean radial 
direction, appears very indirectly in Matrix Theory.  Rather 
than imposing the constraints directly on a a few $X$ fields 
corresponding to the compactified directions, one has to 
scale all the $X$ matrices as well as the worldline time coordinate. 
This approach appears non-trivial and very different 
from  proposals made for a Matrix Model 
for de Sitter made  so far \cite{li}, \cite{chla}, 
and is an interesting avenue for the future.

\bigskip
\centerline{\large \bf Acknowledgements}
\noindent
We are happy to acknowledge useful discussions with Steve Corley,
Atish Dabholkar,  Laurent Freidel,  Aki Hashimoto,  Antal Jevicki, 
Robert de Mello Koch, Joao Rodrigues.   
This research was  supported in part by DOE
grant DE-FE0291ER40688-Task A.

\newpage

{\Large\bf{Appendix A. Limits}}

\renewcommand{\theequation}{A.\arabic{equation}}
\setcounter{equation}{0}

In this appendix we review the various Matrix theory limits, 
and derive the correct limit in our case. 
For completeness, let us recall the formulas relating the M theory
parameters on a spatial circle to the IIA string theory parameters. 
They are obtained from
\be
\frac{1}{g_s l_s}=\frac{1}{R_{11}}~,\;\;\;
\frac{1}{l_s^2}=\frac{R_{11}}{l_P^3}~.
\ee
Sen \cite{sen} and Seiberg \cite{seiberg}
used a construction for M theory compactified on 
$T^p $ in a limit of vanishing radii. We refer to this as $\bar M$
theory, taking 
\be
\bar l_P, \bar R_{11}, \bar R_i \rightarrow 0 \Rightarrow \bar g_s,
\bar l_s \rightarrow 0
\ee
such that 
\be
a_i = \frac{\bar R_i}{\bar l_P}, \;\;\; M=\frac{\bar R_{11}}{\bar l_P^2}
\ee
are held fixed. After dimensionally reducing on $\bar R_{11}$ to string 
theory and making T dualities on all the $\bar R_i$, the T dual variables are
\bea
\tilde{l}_s &=& \bar l_s = M^{-1} \bar g_s^{1/3}\nonumber\\
\tilde{R}_i &=& \frac{\bar l_s^2}{\bar R_i} = \frac{1}{Ma_i}\nonumber\\
\tilde{g}_s &=&\frac{\bar g_s}{\prod_{i=1}^p (\bar R_i/\bar l_s)} =
\bar g_s^{1-p/3}
\prod _{i=1}^p a_i ^{-1}
\eea
and moreover 
\be
\frac{1}{g_{YM}^2}=\frac{\bar l_s^3}{\bar g_s}, \;\; \frac{1}{\tilde{g}_{YM}^2}
=\frac{\bar l_s^{3-p}}{\tilde{g}_s}
\ee
such that 
\be
g_{YM}^2 = M^3, \;\;\; \tilde{g}_{YM}^2 = M^{3-p} \prod_{i=1}^p
a_i^{-1}~.
\ee
So the limit was chosen to decouple string theory both in the original 
and in the dual theory ($\bar g_s, \bar l_s, \tilde{g}_s, \tilde {l}_s
\rightarrow 0$), 
while keeping the Yang-Mills couplings ($g_{YM}$ of the 
D0-branes and $\tilde{g}_{YM}$ of the Dn-branes)  and the dual radii
finite.

Let us now review the BFSS point of view, which
is also  advocated for the Matrix string of type IIA, and the Matrix
theory of type IIB, and then apply it for our case. After that, we
will look at the relation between Sen-Seiberg and BFSS and apply it
to the Matrix models, finally deriving our limit.

BFSS \cite{bfss} chose the limit 
\be
R_{11}\sim N \rightarrow \infty, ~l_P ={\rm fixed}~.
\ee
 
So that 
\be
g_s = (R_{11}/l_P)^{3/2}\rightarrow \infty, \;\;\; l_s
=\frac{l_P^{3/2}}{R_{11}^{1/2}} \rightarrow 0
\ee
and thus one obtained an $U(\infty )$ D0 brane theory. The argument
being that in the limit, string theory does decouple (even though
$g_s$ is infinite), because there are no string states other than D0
branes which have momentum on the 11th direction (that is the D0 
charge), and so if we look at fixed momentum $N/R_{11}$, strings
decouple.

The analogous statement happened for the IIA Matrix string
\cite{motl,bs,dvv}. One added
to the above construction a compactification on a finite $R_9$, and
then made a 9-11 flip, meaning one reinterprets 9 as the 11-th
direction. Since $l_P$ was finite, after the flip
\be
\tilde{g}_s =(\frac{R_9}{l_P})^{3/2}={\rm finite},~ l_s =
\frac{l_P^{3/2}}{R_9^{1/2}}={\rm finite}~.
\ee

The IIB Matrix theory \cite{bs} was similar. Add to the BFSS two extra radii 
$R_1, R_2 \rightarrow 0$, with $R_1/R_2$=finite. We know that M theory 
on this space gives IIB with finite coupling. Then take the BFSS
construction and consider $l_P \rightarrow 0$, but independent of $N$
(which is consistent with the BFSS limit), such that one holds 
$R_1/l_P^3$ and $R_2/l_P^3$ fixed (the $(p,q)$ type IIB string tensions
fixed), then flip 9-11. 

Then 
\be
g_s^B=\frac{R_1}{R_2} ,\;\;\;\; l_s^2=\frac{l_P^3}{R_1}
\ee
are fixed in this limit. 

Similarly for our case, for the new Matrix theory of type IIA obtained
by compactifying on a radius of zero size, we have, first for the
massless case: compactify on an extra $R_3 \rightarrow 0$ and make a T
duality so that 
\be
g_s^A=\frac{g_s^Bl_s}{R_3}=\frac{R_1l_s}{R_2R_3}={\rm fixed}, \;\;\;
l_s^2 =\frac{l_P^3}{R_9}={\rm fixed}
\ee
where we have as before flipped 9-11 (so this corresponds to
IIA with $R_{11}=N$), and we note that now $g_s^B$ goes to
zero, and it is $g_s^A$ which is finite. 
For the massive IIA case, everything is similar, with the
addition of the new parameter $m$. 

The  equivalence of the BFSS limit and the Seiberg-Sen limit \cite{seiberg,sen}
was derived as follows.
The light-like circle compactification for finite $N$ (DLCQ, see 
\cite{susskind})
with $p^+=N/R$ finite (BFSS corresponds to $N$, $R$ infinite, keeping 
$p^+$ finite, with other possible compactified directions of fixed
radii $R_i$), 
\be
\begin{pmatrix} x\\t \end{pmatrix} \sim \begin{pmatrix}
  x\\t\end{pmatrix} +
 \begin{pmatrix} R/\sqrt{2} \\
  -R/\sqrt{2} 
\end{pmatrix}
\ee
is understood as the $R_s \ll R$ limit of 
\be
\begin{pmatrix} x\\t \end{pmatrix} \sim \begin{pmatrix}
  x\\t\end{pmatrix} +
 \begin{pmatrix}
 \sqrt{R^2/2 +R_s^2}  \\
 -R/\sqrt{2} 
\end{pmatrix}
\ee
which is the infinite boost limit ( $\beta=R/\sqrt{R^2+2 R_s^2}$ ) of 
\be
\begin{pmatrix} x\\t \end{pmatrix} \sim \begin{pmatrix}
  x\\t\end{pmatrix} +
 \begin{pmatrix} R_s \\0
\end{pmatrix}~.
\ee

So the light-like compactification of M theory on $R$ is related to the
$R_s \rightarrow 0$ limit of the spatial compactification of another M
theory. If we subsequently rescale Planck's constant such that $p^-$
and $p^i$ are held fixed we obtain the $\bar{M}$ theory described above.
So
\bea
&&\bar{g}_s = (\bar R_{11} \bar{M}_P)^{3/2}, \;\;\; \bar{M}_s^2 = \bar
R_{11}
\bar{M}_P^3\nonumber\\
&& p^+=N/R, \;\; p^-\sim RM_P^2 , \;\; p^i\sim R_i M_P\nonumber\\ 
&&\bar{p_{11}}= N/\bar R_{11} , \;\;
\bar{p}^-\sim \bar R_{11}\bar{M}_P^2, \;\; \bar{p}^i\sim \bar{R}_i \bar{M}_P
\eea
and then 
\be
\bar R_{11} \bar{M}_P^2 =R M_P^2,\;\;\; \bar{R}_i \bar{M}_P= R_i M_P
\ee
are held fixed in the $\bar{M}_P\rightarrow \infty$ limit. Then
\be
\bar{g}_s = (\bar R_{11} \bar{M}_P)^{3/2} = \bar R_{11}^{3/4}
(RM_P^2)^{3/4}\rightarrow 0, ~~\bar{M}_s^2=\bar R_{11} \bar{M}_P^3
=\bar R_{11}^{-1/2}
(RM_P^2)^{3/2}\rightarrow \infty
\ee
so string theory decouples and the D0 coupling is fixed
\be
g_{YM}^2 =\bar{g}_s \bar{M}_s^3= (\bar R_{11} \bar M_P^2)^3= (RM_P^2)^3~.
\ee

If the (BFSS) M theory is compactified on a torus of fixed radii $R_i$
then one T dualizes the string theory coming from the $\bar{M}$ theory
and gets
\bea
\tilde{R}_i &=& \frac{1}{\bar{R}_i \bar{M}_s^2}= \frac{1}{\bar{R}_i
\bar{M}_P^3\bar R_{11}}=\frac{1}{R_i R M_P^3}\nonumber\\
\tilde{g}_s &=& \bar{g}_s \bar{M}_s^p \prod_i \tilde{R}_i= 
\bar{M}_s^{p-3}R^3 M_P^6 \prod_i \tilde{R}_i \rightarrow 0~ {\rm if} 
\;\;\;p<3 \nonumber\\
\tilde{g}_{YM, Dp}^2 &=& \frac{\tilde{g}_s}{\bar{M}_s^{p-3}}= R^3
M_P^6 \prod_i \tilde{R}_i~.
\eea
So again string theory decouples and one gets a Dp brane theory of
fixed Yang-Mills coupling and dual radii. 

Applying this to the Matrix string theory, one relates again M theory
on light-like $R\sim N $, with $l_P$ fixed and $R_9$ fixed to the
$\bar{M}$ theory with $R_s\rightarrow 0, \bar{M}_P\rightarrow 
\infty , \bar{R}_9\rightarrow 0$, that is a D0 brane theory on a
vanishing circle. After T duality, it becomes a D1 brane theory with
fixed $\tilde{R}_9$. 

On the M theory side, one flips 9-11, reinterpreting it as string
theory with a light-like coordinate $R$, so
\be
g_s= (R_9 M_P)^{3/2} , \;\;\; l_s =l_P^{3/2} R_9^{-1/2}, R \sim N l_s~.
\ee
Then on the $\bar{M}$ theory side, string theory decouples:
\bea
\tilde{g}_s&=& R_s^{1/2} (RM_P^2)^{3/2} \tilde{R}_9 \rightarrow
0\nonumber\\
\tilde{M}_s^2 &=& R_s^{-1/2} (RM_P^2)^{3/2}\rightarrow \infty
\eea
and the Yang-Mills parameters are
\be
\tilde{R}_9= \frac{l_s^2}{R}, \;\;\; \tilde{g}_{YM}^2 =\frac{1}{g_s^2 
\tilde{R}_9^2}
\ee
In this way a D1 theory with fixed parameters is related to a string 
theory with  fixed parameters.

For our IIA Matrix model, M theory with $R_i\rightarrow 0,
l_P\rightarrow 0, R \sim N l_s$ with 
\be
g_s^A=\frac{R_1 l_s}{R_2 R_3}, ~l_s=l_P^{3/2} R_1^{-1/2}
\ee
fixed. Passing to the $\bar M$-theory we see we are left with a
decoupled D3-brane theory on the T-dual space with parameters
\be
\tilde{R}_1 =\frac{l_s^2}{R},~ \tilde{R}_2 =\frac{g_s^A l_s R_3}{R},~
\tilde{R}_3=\frac{g_s^A l_s R_2}{R}, ~\tilde{g}_s=\tilde{g}_{YM}^2
=\frac{l_sg_s^A}{R_1}~.
\ee

We notice though that the Yang-Mills coupling and 2 of the radii  are
actually not finite, so there is probably a better description, but one
has to find it. In particular, since the Yang-Mills coupling is going
to infinity, one should S dualize, but the problem is in the presence
of the noncommutativity it is not very obvious what that means, so we 
will stick with this description. 
At $\theta=0$ the S-dual is a good description, and 
\be
\tilde{g}_{s,D}=\tilde{g}^2_{YM, D}=\frac{R_1}{l_s g_s^A}\rightarrow
0~.
\ee
But under S duality, the ``dimensionless Newton constant'' $\bar{k}^2/R^8\sim 
\tilde{g}_s^2(\frac{\bar{l}_s}{R})^8$  (with R a fixed length scale in
the metric) is invariant (since $g_s\rightarrow 1/g_s, R\rightarrow
R/\sqrt{g_s}$ and we would like to have $\bar{l}_s/R\rightarrow 0$ as 
well as $\bar{l}_s/R_D\rightarrow 0$. (then, the S dual theory is  
decoupled, and therefore so is the original theory)
The condition can be written as  
$g_s (\bar{l}_s/R)^2\rightarrow 0$, that is, the coefficient of the
first loop  correction to the action should be
negligible, and this condition is satisfied if we 
have an $\bar{M}$ theory.


We also notice that $\tilde{R}_1$ fixed, but $\tilde{R}_{2,3}
\rightarrow 0$, but all $\tilde{R}_i/\bar{l}_s\rightarrow \infty$ (so 
we are talking about a 3+1d \SYM !) and there are 2 fixed quantities, a
dimensionless one,
$\frac{\tilde{R}_2\tilde{R}_3}{\tilde{R}_1^2}\tilde{g}^2_{YM}$, and a 
dimensionful one, 
$\tilde{R}_1$, which will be related to $g_s^A$ and $l_s^A$,
respectively. 







\bibliography{massive2a.bib}
\bibliographystyle{utphys}

\end{document}